\documentclass[aps,amsfonts,prd,showpacs,nofootinbib,twocolumn,superscriptaddress]{revtex4}
\usepackage{graphicx}
\usepackage{amssymb}

\newcommand {\eq}{\begin{equation}}
\newcommand {\ee}{\end{equation}}

\begin{document}

\title{Dynamics of holographic vacuum energy in the DGP model}

\author{Xing Wu}
\affiliation {Department of
Astronomy, Beijing Normal University, Beijing 100875, People's
Republic of China}

\author{Rong-Gen Cai}
\affiliation{Institute of Theoretical
Physics, Chinese Academy of Sciences, P.O. Box 2735, Beijing 100080,
People's Republic of China}

\author{Zong-Hong Zhu}
\email{zhuzh@bnu.edu.cn}
\affiliation{Department of Astronomy,
Beijing Normal University, Beijing 100875, People's Republic of
China}

\begin{abstract}
We consider the evolution of the vacuum energy in the DGP model
according to the holographic principle under the assumption that the
relation linking the IR and UV cut-offs still holds in this
scenario. The model is studied when the IR cut-off is chosen to be
the Hubble scale $H^{-1}$, the particle horizon $R_{\rm ph}$ and the
future event horizon $R_{\rm eh}$, respectively. And the two
branches of the DGP model are also taken into account. Through
numerical analysis, we find that in the cases of $H^{-1}$ in the (+)
branch and $R_{\rm eh}$ in both branches, the vacuum energy can play
the role of dark energy.  Moreover, when considering the combination
of the vacuum energy and the $5D$ gravity effect in both branches,
the equation of state of the effective dark energy may cross $-1$,
which may lead to the Big Rip singularity. Besides, we constrain the
model with the Type Ia supernovae and baryon oscillation data and
find that our model is consistent with current data within
$1\sigma$, and that the observations prefer either a pure
holographic dark energy or a pure DGP model.
\end{abstract}

\maketitle
\section{Introduction}

Recent SNe Ia and WMAP observations \cite{SN,WMAP} have indicated
that our universe is currently undergoing an accelerating expansion,
which confront the fundamental theories with great challenges and
also make the researches on this problem a major endeavor in modern
astrophysics and cosmology. The origin of the cosmic acceleration is
still a mystery and is referred to as the dark energy problem.
Various models have been proposed to solve this problem. They
generally fall into the following two ways. One is to add an exotic
energy component with negative pressure, that is, the dark energy,
to the total energy budget of the universe. Among others, the most
competitive candidate of dark energy sofar is the cosmological
constant due to both its theoretical simpleness and its great
success in fitting with observational data, although it suffers from
the cosmological constant problem\cite{cc} . Such problem is
expected to be solved or alleviated in the models of dynamic dark
energy (see \cite{0603057} for a more detailed review), which
generally contains a scalar field evolving in time and driving the
acceleration, just like the scalar field introduced for the
inflation stage at early universe. In fact, the cosmological
constant problem is essentially a problem of quantum gravity. In
quantum field theory, where the effect of gravity is neglected, the
vacuum energy is determined by the UV cut-off $k_c$, that is
$\rho_{\Lambda}\propto k_c^4$. No matter how we choose the UV
cut-off, be it the Planck scale ~$10^{19}$GeV or the electroweak
scale ~TeV, the value predicted by theory is far greater that that
observed ~$10^{-47}\rm GeV^4$. Since we are concerning problems at
the cosmological scale, however, we should have take into account
the effect of gravity. It is expected that the  value of the
cosmological constant or the vacuum energy would be predicted
correctly from a complete theory of quantum gravity, which is still
being explored. But at present the holographic principle, which is
believed to be an important feature of quantum gravity, may shed
some light on solving this problem. Follow the line of the
holographic principle, the holographic dark energy model\cite{Li} is
a promising candidate for solving the dark energy problem. In this
model, the vacuum energy is no longer a time-independent constant,
but evolves with time according to the holographic principle. The
vacuum energy is related to the length measure on cosmological scale
\eq \rho_{\Lambda}=\frac{3c^2M_p^2}{L^2} \label{holo} ,\ee where
$\rho_{\Lambda}$ is directly related to the UV cut-off, $L$ is the
IR cut-off, $M_p$ denotes the Planck mass and $c$ is a numerical
factor by convention, which is the parameter of the model. This
relation of the entanglement of UV/IR was first proposed in
\cite{cohen}, where $L$ was first chosen as the Hubble scale. Then
Hsu\cite{Hsu} pointed out that this would not lead to the desired
equation of state. Finally, Li\cite{Li} proposed the holographic
dark energy model where $L$ is the event horizon. And this model
fits very well with current observations\cite{holo fit}.

The other way to solve the dark energy problem is to modify the
theory of gravity at large scale, without resorting to any new
energy component. For example, the $f(R)$ theory\cite{f(R)} modifies
the standard Einstein-Hilbert action to introduce an effective dark
energy component in the Einstein frame. Here we focus on the DGP
model \cite{DGP}, which describes our universe as a $4D$ brane
embedded in a $5D$ Minkowski bulk and explains the origin of the
dark energy as the gravity on the brane leaking into the bulk at
large scale. The model is described by the action

\begin{eqnarray}
S &=& -\frac{M^3_{(5)}}{2}\int
d^5X\sqrt{-g}R_{(5)}-\frac{M_p^2}{2}\int d^4x\sqrt{-h}R_{(4)}
\nonumber \\ &+& \int d^4x\sqrt{-h}\mathcal{L}_m+S_{GH} ,
\end{eqnarray}
where $g_{ab}$ is the bulk metric and $h_{\mu\nu}$ is the induced
metric on the brane. The first term contains the $5D$ Ricci scalar
whereas the second term contains the $4D$ Ricci scalar on the brane,
which is an extra term due to quantum effects, in contrast to the
Randall-Sundrum scenario\cite{RS}. The third term represents matter
localized on the brane. And $S_{GH}$ is the Gibbsons-Hawking
boundary term.

In this paper, we assume that the relation Eq.(\ref{holo}) still
holds in the DGP model, and consider the evolution of the vacuum
energy on the brane (or the brane tension) according to the
holographic principle. Note that there are other models also
generalizing the standard DGP model by adding a cosmological
constant(LDGP)\cite{LDGP}, a Quiessence perfect
fluid(QDGP)\cite{QDGP}, a scalar field(SDGP)\cite{SDGP}, or the
Chaplygin gas(CDGP)\cite{CDGP}. Although the holographic dark energy
model is well consistent with observational data, it should be noted
that the core of the holographic principle is that it relates the UV
and IR cut-offs of a local quantum field system, which reflects some
feature of quantum gravity. Thus even if the holographic vacuum
energy had not played the role of dark energy, it would still be of
significance to study on this problem.

The paper is organized as follows. In section \ref{sec2}, we present
the model under three cases of choosing the IR cut-off for the
vacuum energy as respectively the Hubble scale $H^{-1}$, the
particle horizon $R_{\rm ph}$ and the future event horizon $R_{\rm
eh}$. We study the equation of state (EoS) of the vacuum energy and
the effective EoS due to the combined effect of both the holographic
vacuum energy and the 5D gravity effect. In section \ref{sec3}, we
use recent observational data to constrain the model and fit its
parameters. We conclude this paper in the final section.

\section{The model}\label{sec2}
We assume a flat, homogeneous and isotropic brane in accordance with
the result of the WMAP observation\cite{WMAP}. Following \cite{DGP},
the Friedmann equation is \eq
H^2=(\sqrt{\frac{\rho}{3M_p^2}+\frac{1}{4r_c^2}}+\epsilon\frac{1}{2r_c})^2
\label{F1} ,\ee or equivalently \eq
H^2-\epsilon\frac{H}{r_c}=\frac{1}{3M_p^2}\rho\label{F2} ,\ee where
$H\equiv\ \dot{a}/a$ is the Hubble parameter and $r_c\equiv
M_p^2/2M^3_{(5)}$ is the distance scale reflecting the competition
between 4D and 5D effects of gravity. For $H^{-1}\ll r_c$ (early
times), the 4D general relativity is recovered; for $H^{-1}\gtrsim
r_c$ (late times), the 5D effect begins to be significant.
$\epsilon=\pm$ represents two branches of the model of which the
$(+)$ branch is the self-accelerating solution in which the universe
may enter into an accelerating phase in late time by virtue of pure
5D effect of gravity, while the $(-)$ branch cannot undergo an
acceleration without additional dark energy component. Here the
vacuum energy is added in $\rho=\rho_m+\rho_{\Lambda}$. Obviously
the dynamics is typically different from the standard FRW cosmology.
In addition, we have the usual equation of conservation \eq
\dot{\rho}+3H(1+w)\rho=0 ,\ee where $w$ is the equation of state
(EoS). Here we assume that there is no interaction between matter
and vacuum energy. Therefore both components obey the equation of
conservation respectively and, in particular, for matter we have
$\rho_m=\rho_{m0}a^{-3}$, the same as that in usual FRW cosmology.
We note that here the EoS of the vacuum energy evolves with time due
to the holographic principle as shown later, as opposed to
$\Lambda$CDM where $w_{\Lambda}\equiv-1$.

When applying the holographic principle in cosmology, a crucial
problem is how to choose the IR cut-off L. In usual FRW sosmology,
it is shown\cite{Li} that only in the case of choosing L as the
future event horizon can vacuum energy play the role of dark energy.
In the DGP framework, we consider L as the Hubble scale $H^{-1}$,
the particle horizon $R_{\rm ph}$ and the future event horizon
$R_{\rm eh}$ respectively. And then we investigate the evolution of
the EoS of the vacuum energy.

Intuitionally, we may guess that for one thing, in the (-) branch
where there is no self-acceleration, adding a component whose ${\rm
EoS}>-1/3$ will not lead to an acceleration whereas adding a dark
energy component may cause the universe in this branch to
accelerate; for the other thing, in the (+) branch, no matter
whether the holographic vacuum energy we add can in itself play the
role of dark energy, the combined effect may lead to an acceleration
due to the contribution from the self-acceleration of this branch.

\subsection{L as $H^{-1}$}
By Eq.(\ref{holo}), the vacuum energy is
$\rho_{\Lambda}=3c^2M_p^2H^2$. For convenience we insert this result
into the Friedmann Eq.(\ref{F2}). After defining
$\Omega_m=\frac{\rho_m}{3M_p^2H_0^2}=\Omega_{m0}(1+z)^3$,
$\Omega_{\Lambda}=\frac{\rho_{\Lambda}}{3M_p^2H_0^2}$,
$\Omega_r=\frac{1}{4r_c^2H_0^2}$ and
$\Omega_{\Lambda}=\frac{\rho_{\Lambda}}{3M_p^2H_0^2}=c^2\frac{H^2}{H_0^2}$,
where $H_0$ is the Hubble parameter at redshift $z=0$, the above
equation can be transformed into \eq
(1-c^2)(\frac{H}{H_0})^2-2\epsilon\sqrt{\Omega_{r_c}}\frac{H}{H_0}-\Omega_m=0\label{Hubble
equation} ,\ee Then we can solve the above equation to get
$E=H/H_0$.

\begin{itemize}

\item
{Case 1: $c=1$} \eq -\epsilon\frac{H}{r_c}=\frac{1}{3M_p^2}\rho_m
,\ee where $\epsilon=+1$ corresponds to the contracting solution,
and $\epsilon=-1$ represents the expanding solution \eq
H=\frac{r_c}{3M_p^2}\rho_m .\ee In this case, the solution is easy
to find by virtue of the conservation equation of $\rho_m$:
$\rho_m\propto a^{-3}$ therefore $\rho_{\Lambda}\propto H^2\propto
\rho^2_m\propto a^{-6}$, namely, the vacuum energy decreases faster
than $\rho_m$ and it cannot be dominant in late time. Thus we do not
consider this case as of physical interest.

When $c\neq1$ we write the general solution \eq
\frac{H}{H_0}=\frac{\epsilon\sqrt{\Omega_{r_c}}\pm\sqrt{\Omega_{r_c}+\Omega_m(1-c^2)}}{1-c^2}\label{friedmann1}
.\ee

\item{
Case 2: $c>1$} Since $\Omega_{r_c}$ is constant, the part under the
square root on the RHS of Eq.(\ref{friedmann1}) may become less than
zero as z increases, therefore the solution in this case is
unphysical.

\item{
Case 3: $c<1$} The physical solution is \eq
\frac{H}{H_0}=\frac{\epsilon\sqrt{\Omega_{r_c}}+\sqrt{\Omega_{r_c}+\Omega_m(1-c^2)}}{1-c^2}
.\ee The parameters in this expression should satisfy the following
condition according to Eq.(\ref{Hubble equation}) at $z=0$ \eq
1-c^2-2\epsilon\sqrt{\Omega_{r_c}}-\Omega_{m0}=0\label{initial_H}
.\ee

\end{itemize}

Thereby we obtain $\Omega_{\Lambda}$ and its derivative with respect
to z in the following forms
\begin{eqnarray}
\Omega_{\Lambda} &=&
\frac{c^2}{(1-c^2)^2}[\Omega_m(1-c^2)+2\Omega_{r_c} \nonumber
\\ &+& 2\epsilon\sqrt{\Omega_{r_c}}\sqrt{\Omega_{r_c}+\Omega_m(1-c^2)}]\label{omega_L_H},
\end{eqnarray}
\begin{eqnarray} \Omega'_{\Lambda} &=& \frac{c^2}{H_0^2}2HH' \nonumber \\
&=&
c^2\frac{\epsilon\sqrt{\Omega_{r_c}}+\sqrt{\Omega_{r_c}+\Omega_m(1-c^2)}}{1-c^2}\nonumber
\\ &\times& \frac{3\Omega_m(1+z)^{-1}}{\sqrt{\Omega_{r_c}+\Omega_m(1-c^2)}}\label{domega_L_H}.
\end{eqnarray}
Furthermore, we require that matter dominate over vacuum energy as z
grows larger, or else it would spoil the success of standard Big
Bang cosmology. From Eq.(\ref{omega_L_H}) we obtain the asymptotic
expression of $\Omega_{\Lambda}$ for large $z$ \eq
\Omega_{\Lambda}=\frac{c^2}{1-c^2}\Omega_m .\ee Thus if we demand
$\Omega_m$ dominate over $\Omega_{\Lambda}$ at early time, we have
$\frac{c^2}{1-c^2}\ll 1$ or $c\ll 1/\sqrt{2}$. In the following we
only consider the case of $c<1/\sqrt{2}\sim0.7$.

\subsection{L as the particle horizon and the event horizon}
The IR cut-off L is given by the definition of these two horizons
\begin{eqnarray}
L =  \left\{\begin{array}{lll}
R_{\rm ph}=a(t)\int_0^t\frac{dt'}{a(t')}=a\int_0^a\frac{da'}{Ha'^2} \\
\\
R_{\rm
eh}=a(t)\int_t^{\infty}\frac{dt'}{a(t')}=a\int_a^{\infty}\frac{da'}{Ha'^2}
\end{array} \right.
\label{L}\end{eqnarray} Here we use the Friedmann Eq.(\ref{F1}) and
recast it into \eq
E(z)=\frac{H}{H_0}=\sqrt{\Omega_m+\Omega_{\Lambda}+\Omega_{r_c}}+\epsilon\sqrt{\Omega_{r_c}}\label{E}
.\ee

By Eq.(\ref{holo}) we have \eq L=\sqrt{3c^2M_p^2\over
\rho_{\Lambda}} .\ee Inserting Eq.(\ref{L}) and Eq.(\ref{E}) into
the above equation, taking derivative with respect to $a$ on both
sides, and then using $1+z={1 / a}$, we obtain the evolution
equation of $\Omega_{\Lambda}$ with respect to z \footnote{It should
be noted that here we define the fractional energy densities with
$H_0$ rather than with $H$, by which these fractions are often
defined in literature.}
\eq\Omega'_{\Lambda}=(1+z)^{-1}\frac{2}{c}\Omega_{\Lambda}^{3/2}
(\frac{\theta}{\sqrt{\Omega_m+\Omega_{\Lambda}+\Omega_{r_c}}+\epsilon\sqrt{\Omega_{r_c}}}+\frac{c}{\sqrt{\Omega_{\Lambda}}})\label{domegape}
,\ee where $\theta=+1$ corresponds to $L=R_{\rm ph}$ and $\theta=-1$
to $L=R_{\rm eh}$ and the initial condition of this differential
equation is given by setting z=0 in Eq.(\ref{E})
\eq\Omega_{\Lambda0}=1-\epsilon2\sqrt{\Omega_{r_c}}-\Omega_{m0}\label{initial}
.\ee We can solve this equation numerically and require that, as
mentioned above, $\Omega_{\Lambda}$ should become negligible
compared with $\Omega_m$ as z grows.

In fact, it is only in an eternally accelerating universe that the
event horizon exists. Thus, when using the event horizon as the IR
cut-off, we already assume an accelerating universe and therefore
the existence of some effective dark energy.

\subsection{EoS of the vacuum energy}
By energy conservation we have \eq
\dot{\rho}_{\Lambda}+3H(1+w_{\Lambda})\rho_{\Lambda}=0  ,\ee from
which we get \eq
w_{\Lambda}=-1-\frac{\dot{\rho_{\Lambda}}}{3H\rho_{\Lambda}}=-1+(1+z)\frac{\Omega'_{\Lambda}}{3\Omega_{\Lambda}}\label{wL}
,\ee where $1+z=1/a$ is used.

\begin{figure} 
\begin{center}
\includegraphics[scale=0.40]{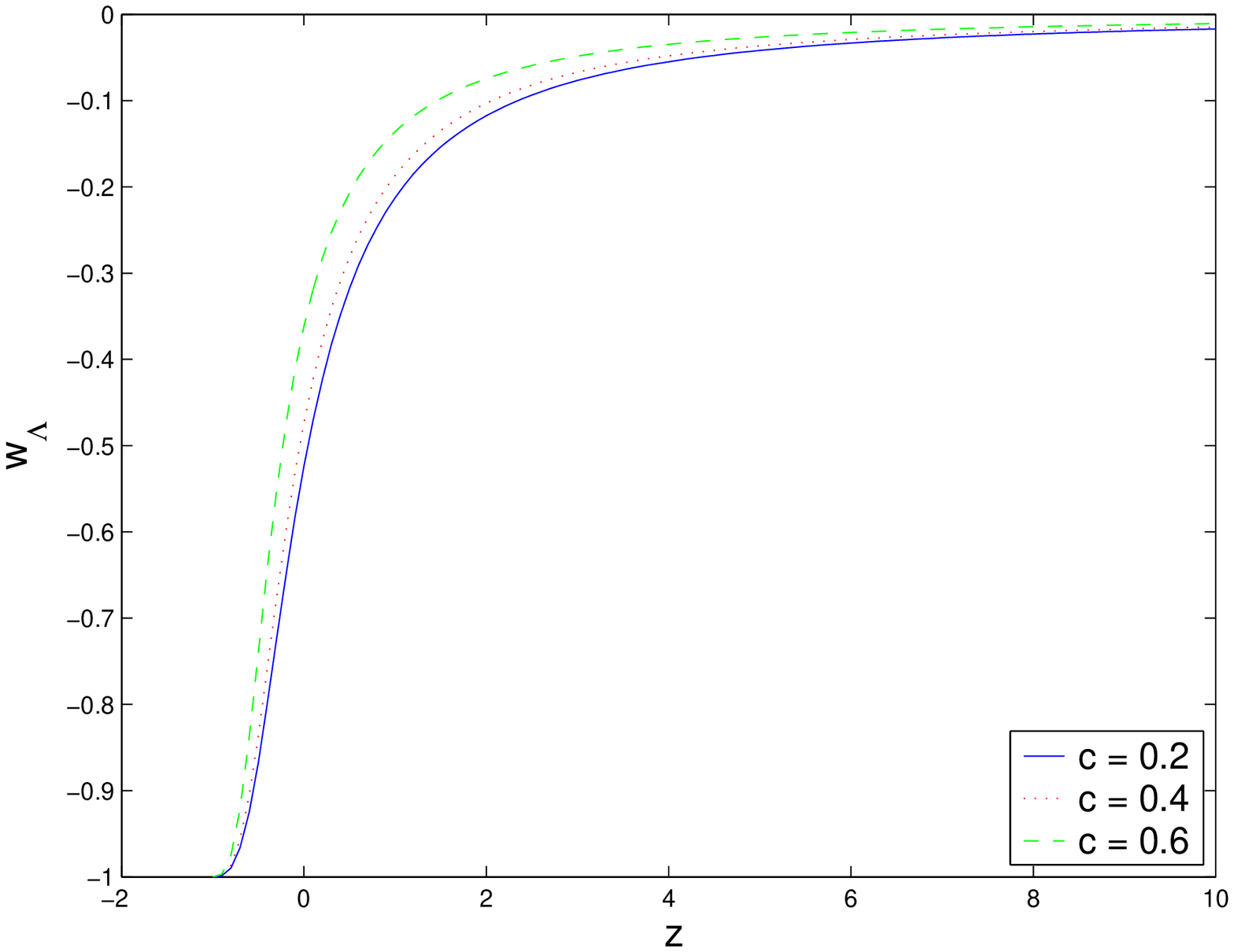}
\includegraphics[scale=0.40]{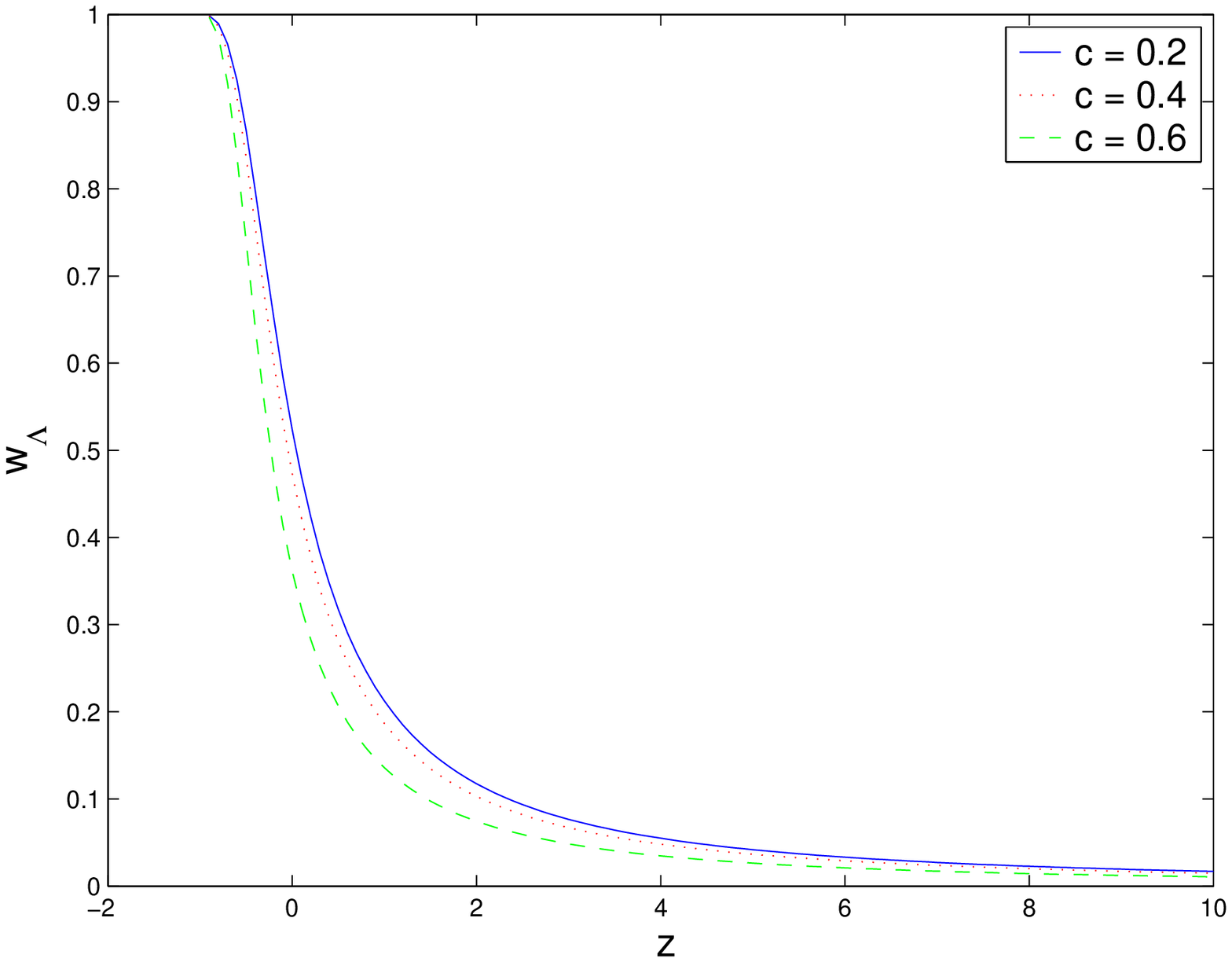}
\caption[]{\small The evolution of $w_{\Lambda}(z)$. $L=H^{-1}$
  and $\Omega_m=0.3$. In
the branch $\epsilon=+1$(upper), the EoS evolves from zero in the
past to $-1$ in the future. It is currently less than $-1/3$ and
therefore $\Omega_{\Lambda}$ can serve as dark energy. In the branch
$\epsilon=-1$(lower), however, the EoS is always positive.}
\label{fig:w_L_H_1}
\end{center}
\end{figure}

For L as Hubble scale, we insert Eq.(\ref{omega_L_H}) and
Eq.(\ref{domega_L_H}) into (\ref{wL}) to get
\begin{eqnarray}
w_{\Lambda} &=&
-1+\frac{1-c^2}{\epsilon\sqrt{\Omega_{r_c}}+\sqrt{\Omega_{r_c}+\Omega_m(1-c^2)}}
\nonumber \\ &\times&
\frac{\Omega_m}{\sqrt{\Omega_{r_c}+\Omega_m(1-c^2)}} ,
\end{eqnarray}
where $c<1$ is required. Some features of the evolution of
$w_{\Lambda}$ can be shown analytically if we rewrite this equation
as \eq w_{\Lambda}=-1+{1\over{\epsilon\sqrt{F}\sqrt{F+1}+F+1}} ,\ee
where $F\equiv {\Omega_{r_c}\over{\Omega_m(1-c^2)}}>0$. For
$\epsilon=+1$, the denominator is always greater than 1, leading to
$w_{\Lambda}<0$ forever; for $\epsilon=-1$, the denominator is
always less than 1, leading to $w_{\Lambda}>0$. The evolution of
$w_{\Lambda}$ is shown numerically in Fig.1, where we fix
$\Omega_{m0}=0.3$ and set different values of $c$. From the figure
we find that in the (-) branch the EoS is always positive, therefore
the vacuum energy can not drive the cosmic acceleration. In the (+)
branch, however, for smaller $c$ ( e.g. $c=0.2,0.4$ on the plot),
$w_{\Lambda}(z=0)$ may become less than $-1/3$ and the vacuum energy
may play the role of dark energy. This is different from the case in
the usual FRW universe, where $H^{-1}$ cannot serve as the IR
cut-off of the holographic dark energy.


\begin{figure} 
\begin{center}
\includegraphics[scale=0.40]{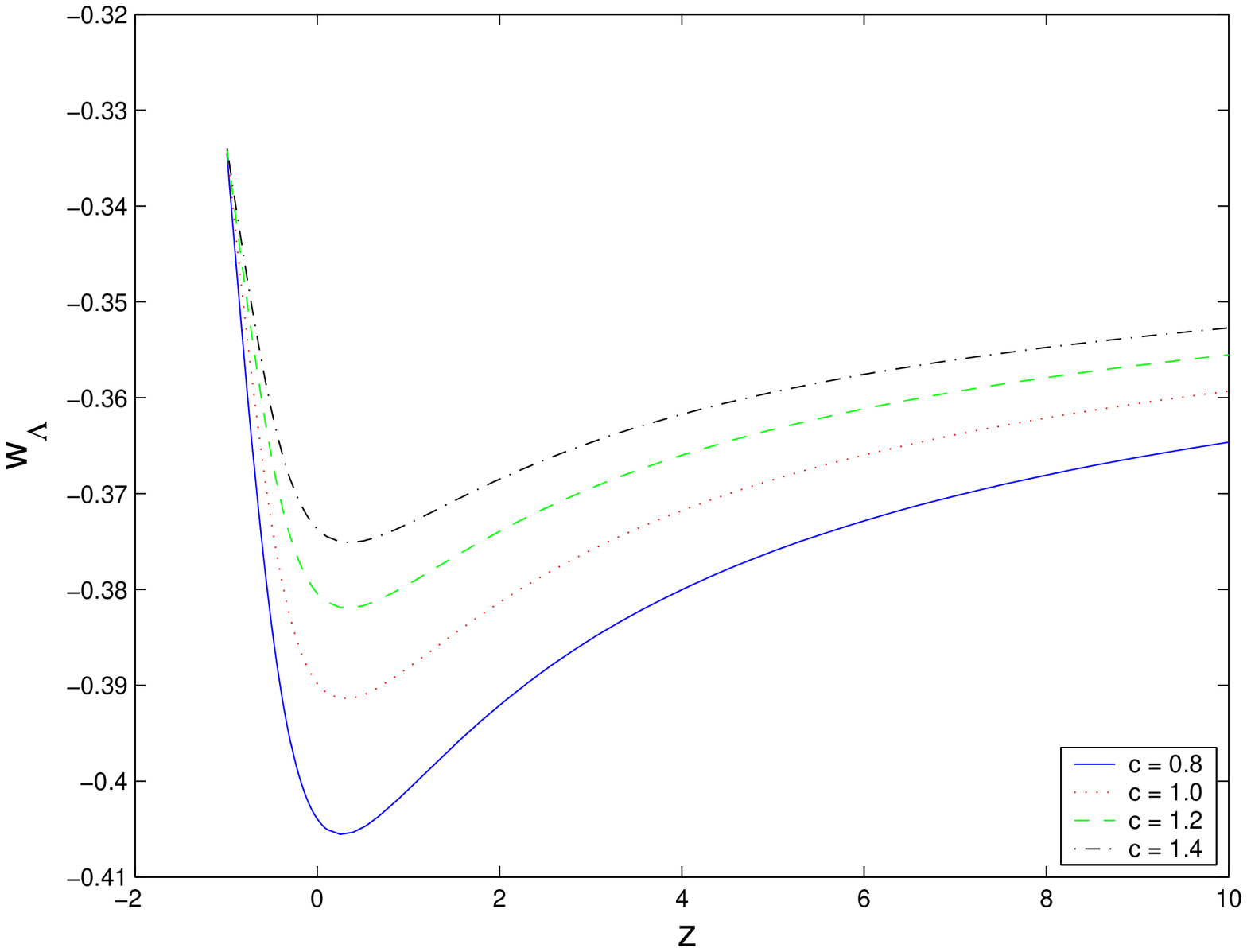}
\includegraphics[scale=0.40]{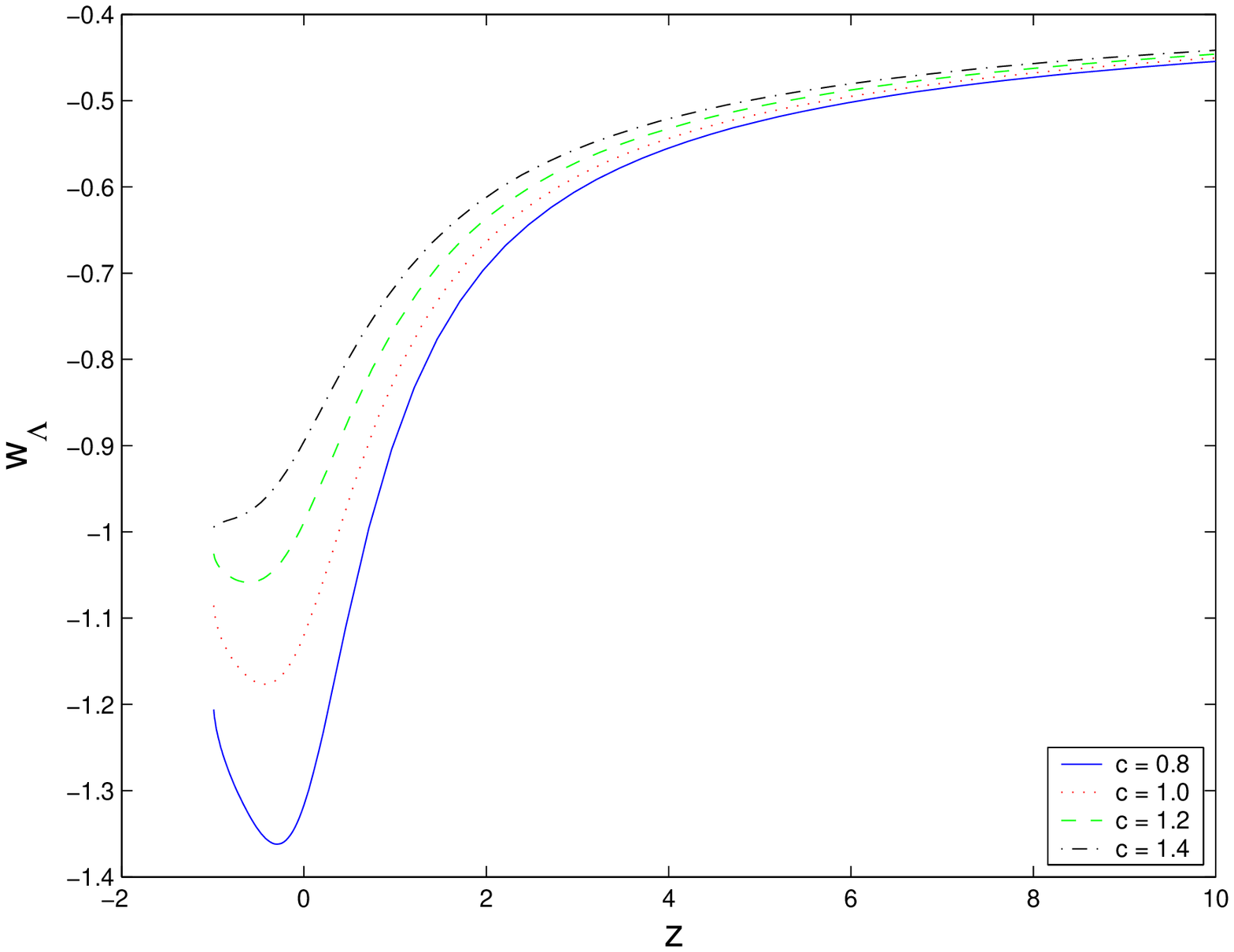}
\caption[]{\small The evolution of $w_{\Lambda}(z)$. $L=R_{\rm eh}$,
  $\Omega_m=0.3$ and
$\Omega_{r_c}=0.12$. In both cases the EoS is always less than
$-1/3$ and the vacuum energy can serve as dark energy. In the branch
$\epsilon=+1$(upper), the EoS ends up with $-1/3$ in the future,
while in the branch $\epsilon=-1$(lower) it may cross $-1$ for some
values of the parameters. }\label{fig:w_L_eh_1}
\end{center}
\end{figure}

For L as $R_{\rm ph}$ and $R_{\rm eh}$ we insert
Eq.({\ref{domegape}) into Eq.(\ref{wL}) to get \eq w_{\Lambda}=
-\frac{1}{3}+\frac{2}{3c}\frac{\theta\sqrt{\Omega_{\Lambda}}}{\sqrt{\Omega_m+\Omega_{\Lambda}+\Omega_{r_c}}+\epsilon\sqrt{\Omega_{r_c}}}
.\ee From this equation we can easily see that, it is only when
$\theta=-1$ that $w_{\Lambda}<-\frac{1}{3}$, namely, the vacuum
energy serves as dark energy if L is $R_{\rm eh}$ rather than
$R_{\rm ph}$. This is the same as in the usual FRW universe.
Fig.\ref{fig:w_L_eh_1} shows the evolution of the EoS for the case
of L as $R_{\rm eh}$.

\subsection{EoS of the effective dark energy}
In order to explore the possibility of realizing accelerating
expansion in our model, a combined effect of both the vacuum energy
and the 5D gravity effect should be considered. That is, we need to
find out the EoS of the effective dark energy. Firstly, we rewrite
Eq.(\ref{E}) as \eq
E^2=\Omega_m+\Omega_{\Lambda}+2\Omega_{r_c}+2\epsilon\sqrt{\Omega_{r_c}}\sqrt{\Omega_m+\Omega_{\Lambda}+\Omega_{r_c}}
,\ee and then compare this expression with the Friedmann equation in
usual 4D FRW cosmology consisting of a matter component and an
effective dark energy \eq E^2=\Omega_m+\Omega_{\rm eff},\ee where,
by the conservation equation for the effective dark energy, we have

\eq \Omega_{\rm eff}=\Omega_{\rm eff}^{(0)} {\rm exp}
(3\int_0^z\frac{1+w_{\rm eff}(z')}{1+z'}dz'), \ee  and $w_{\rm eff}$
is the EoS of the effective dark energy. We find that \eq\Omega_{\rm
eff}=\Omega_{\Lambda}+2\Omega_{r_c}+2\epsilon\sqrt{\Omega_{r_c}}\sqrt{\Omega_m+\Omega_{\Lambda}+\Omega_{r_c}}
.\ee Taking derivative on both sides of the above equation with
respect to $z$, we obtain
\begin{eqnarray}
1&+&w_{\rm eff}
=\frac{1}{3}\frac{1}{\Omega_{\Lambda}+2\Omega_{r_c}+2\epsilon\sqrt{\Omega_{r_c}}\sqrt{\Omega_m+\Omega_{\Lambda}+\Omega_{r_c}}}\nonumber
\\ &\times&\left [\epsilon\sqrt{\Omega_{r_c}}
\frac{3\Omega_m+\Omega'_{\Lambda}(1+z)}{\sqrt{\Omega_m+\Omega_{\Lambda}+\Omega_{r_c}}}+\Omega_{\Lambda}'(1+z)\right
] ,
\end{eqnarray}
with the same constraint as Eq.(\ref{initial}). Here we also require
the asymptotic behavior of $\Omega_{\rm eff}$ should be dominated
over by $\Omega_m$ in the past for the same reason mentioned above.

For L as the Hubble scale, we show the evolution of the effective
EoS in Fig.\ref{fig:w_eff_H_1} and Fig.\ref{fig:w_eff_H_2}. Clearly
in the (+) branch $w_{\rm eff}$ can become less than $-1/3$ and end
up with $w_{\rm eff}=-1$, and so is $w_{\rm total}$ due to the
effective dark energy dominating over matter at late time.  For L as
the particle horizon, Fig.\ref{fig:w_eff_ph_1} and
Fig.\ref{fig:w_eff_ph_2} show that in the (+) branch the effective
dark energy can drive the cosmic acceleration whereas it cannot in
the other branch. Note that, by Fig.\ref{fig:w_eff_ph_1a}, for small
values of $c$, although the vacuum energy may serve as dark energy,
it can dominate over matter as z grows and consequently spoil the
BBN and structure formation. As Fig.\ref{fig:w_eff_eh_1} and
Fig.\ref{fig:w_eff_eh_2} show, for L as the future event horizon in
both branches, $w_{\rm eff}$ as well as $w_{\rm total}$ may become
less than $-1/3$. Therefore in both branches acceleration may occur.
Besides, there are two points worth particular mentioning: $(1)$
From Fig.\ref{fig:w_eff_H_2} and Fig.\ref{fig:w_eff_ph_2} we see
that in the (-) branch, there exist a pole where $w_{\rm eff}$
diverges. This occurs because $\Omega_{\rm eff}$ evolves from
positive to negative (or the opposite) as z decreases, crossing the
point $\Omega_{\rm eff}=0$ at some $z^{*}$, which implies the
breakdown of the effective description, rather than any pathology of
the model. This can also be confirmed from the plot of $w_{\rm
total}$ where the EoS is well behaved. $(2)$ The EoS in
Fig.\ref{fig:w_eff_eh_1a} to Fig.\ref{fig:w_eff_eh_2a} exhibits
phantom behavior. This is because the effective dark energy posses
phantom behavior and it dominates over matter at late time.
Therefore $\Omega_{\rm eff}$ increases with time until the Big Rip
singularity.

\begin{figure}
\begin{center}
\includegraphics[scale=0.40]{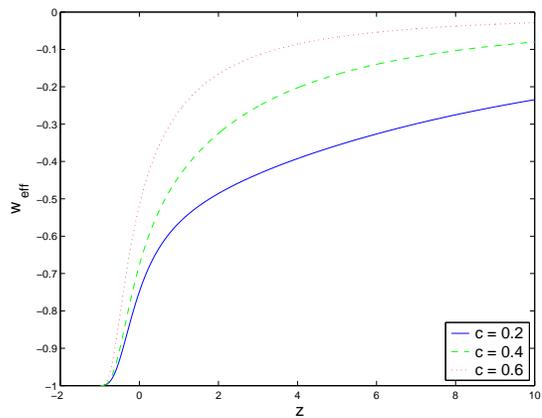}
\caption[]{\small The evolution of $w_{\rm eff}$. $L=H^{-1}$,
$\epsilon=+1$ and $\Omega_m=0.3$. The effective EoS may become less
than $-1/3$ in the near past and end up with $-1$ in the future,
therefore an acceleration may occur in this case.}
\label{fig:w_eff_H_1}
\end{center}
\end{figure}

\begin{figure} 
\begin{center}
\includegraphics[scale=0.4]{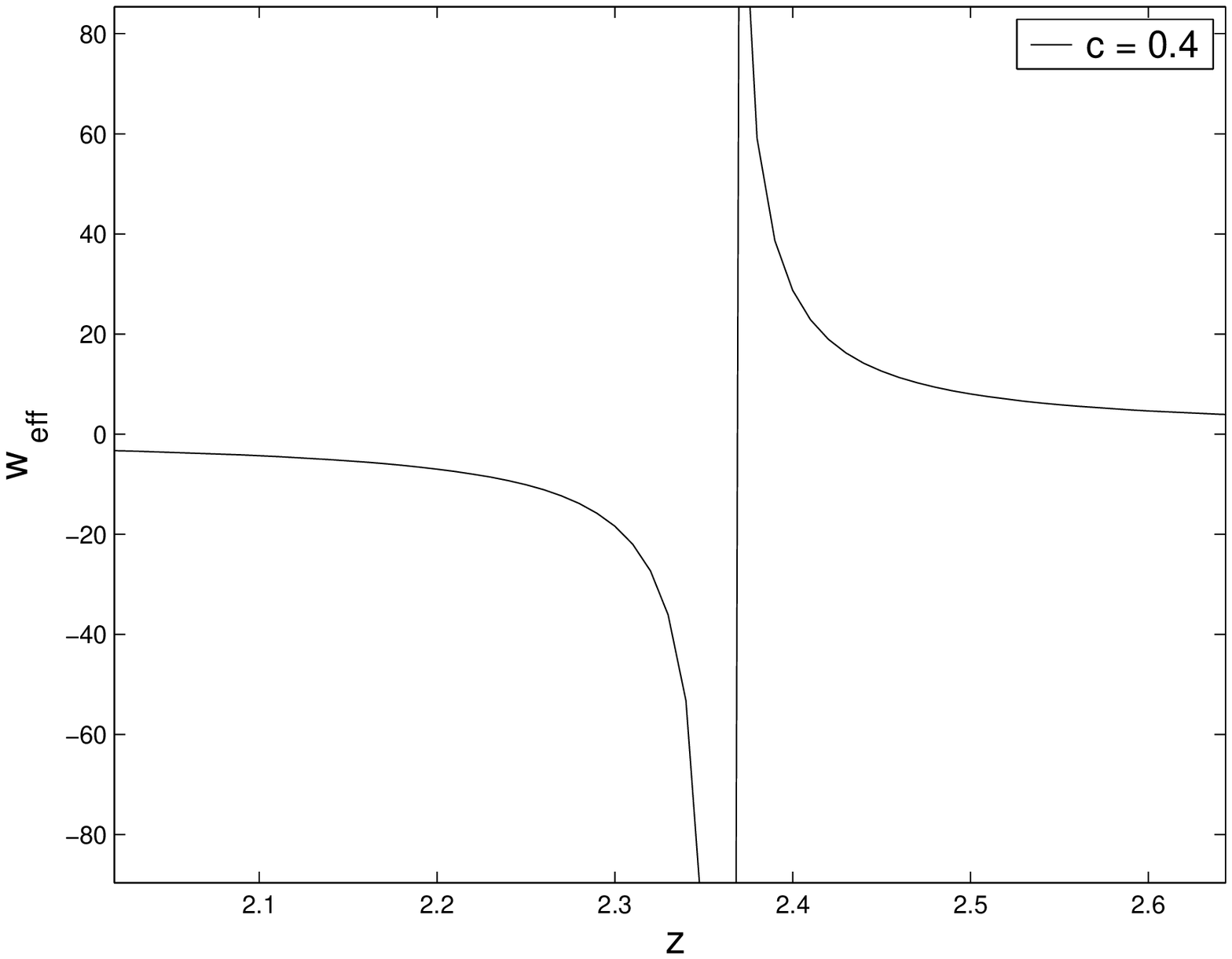}
\includegraphics[scale=0.4]{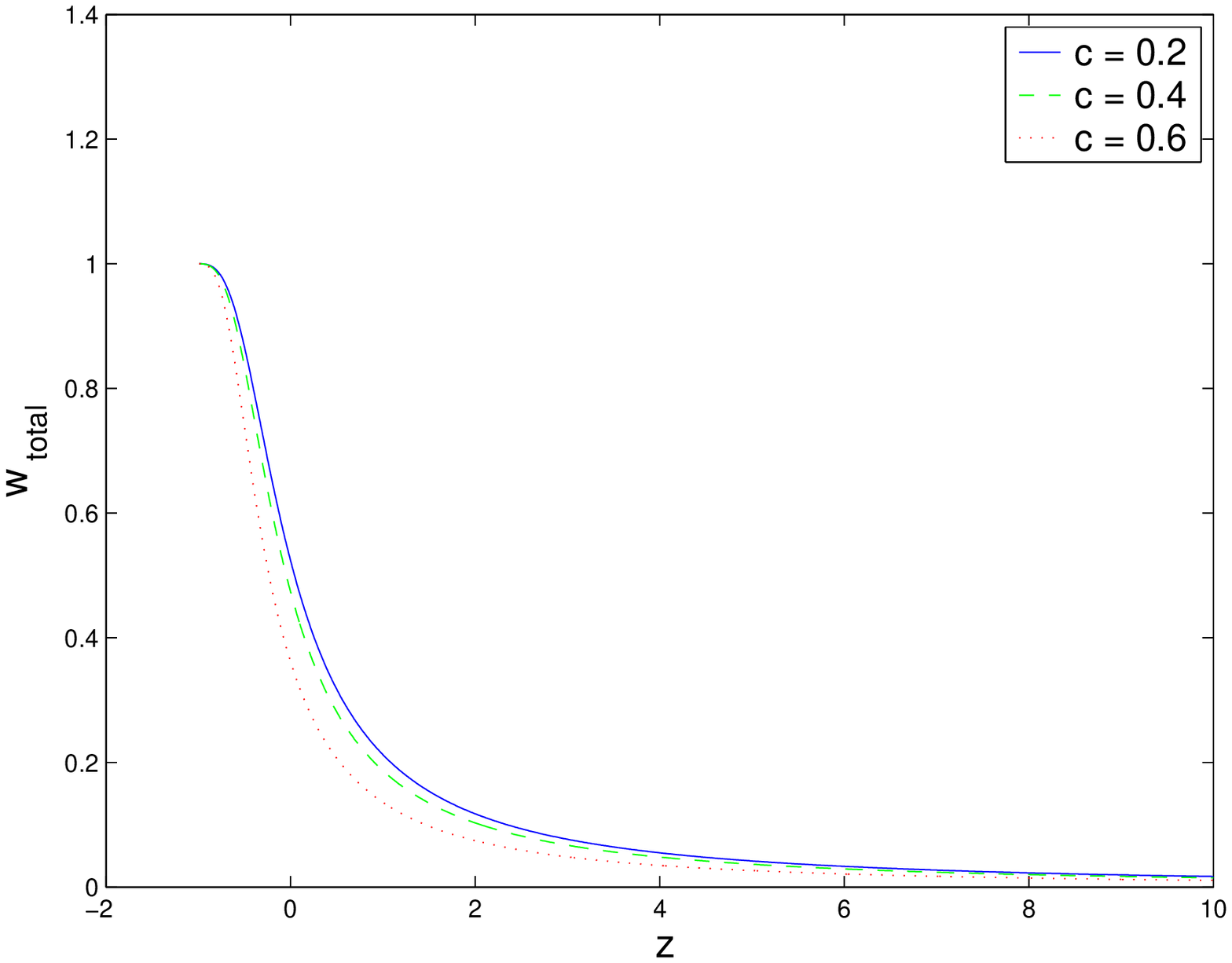}
\caption[]{\small The evolution of $w_{\rm eff}$(upper) and $w_{\rm
total}$(lower). $L=H^{-1}$, $\epsilon=-1$ and $\Omega_m=0.3$. There
is a pole at corresponding redshift $z^{*}$ where $w_{\rm eff}$
becomes divergent. This only means that the effective description
breaks down around $z^{*}$, rather than that some pathology exists
in the model. This can be illustrated by the plot of $w_{\rm
total}$, which is well behaved.}\label{fig:w_eff_H_2}
\end{center}
\end{figure}

\begin{figure}
\begin{center}
\includegraphics[scale=0.40]{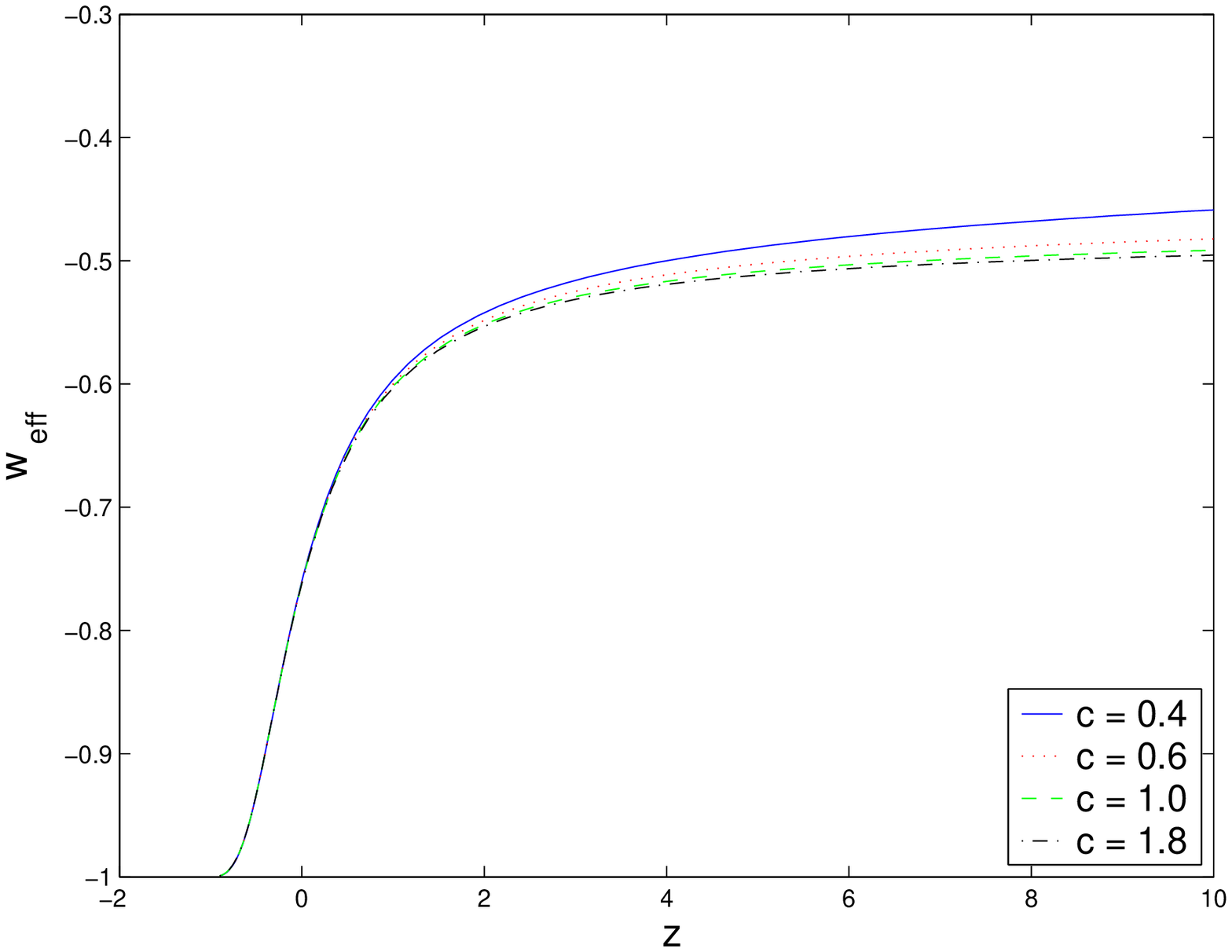}
\caption[]{\small The evolution of $w_{\rm eff}$.  $L=R_{\rm ph}$,
$\epsilon=+1$, $\Omega_m=0.3$ and $\Omega_{r_c}=0.12$. $w_{\rm
eff}(0)<-1/3$ and an acceleration may occur. $w_{\rm eff}\rightarrow
-1$ in the future and the effective dark energy ends up as a
cosmological constant.
  } \label{fig:w_eff_ph_1}
\end{center}
\end{figure}

\begin{figure}
\begin{center}
\includegraphics[scale=0.40]{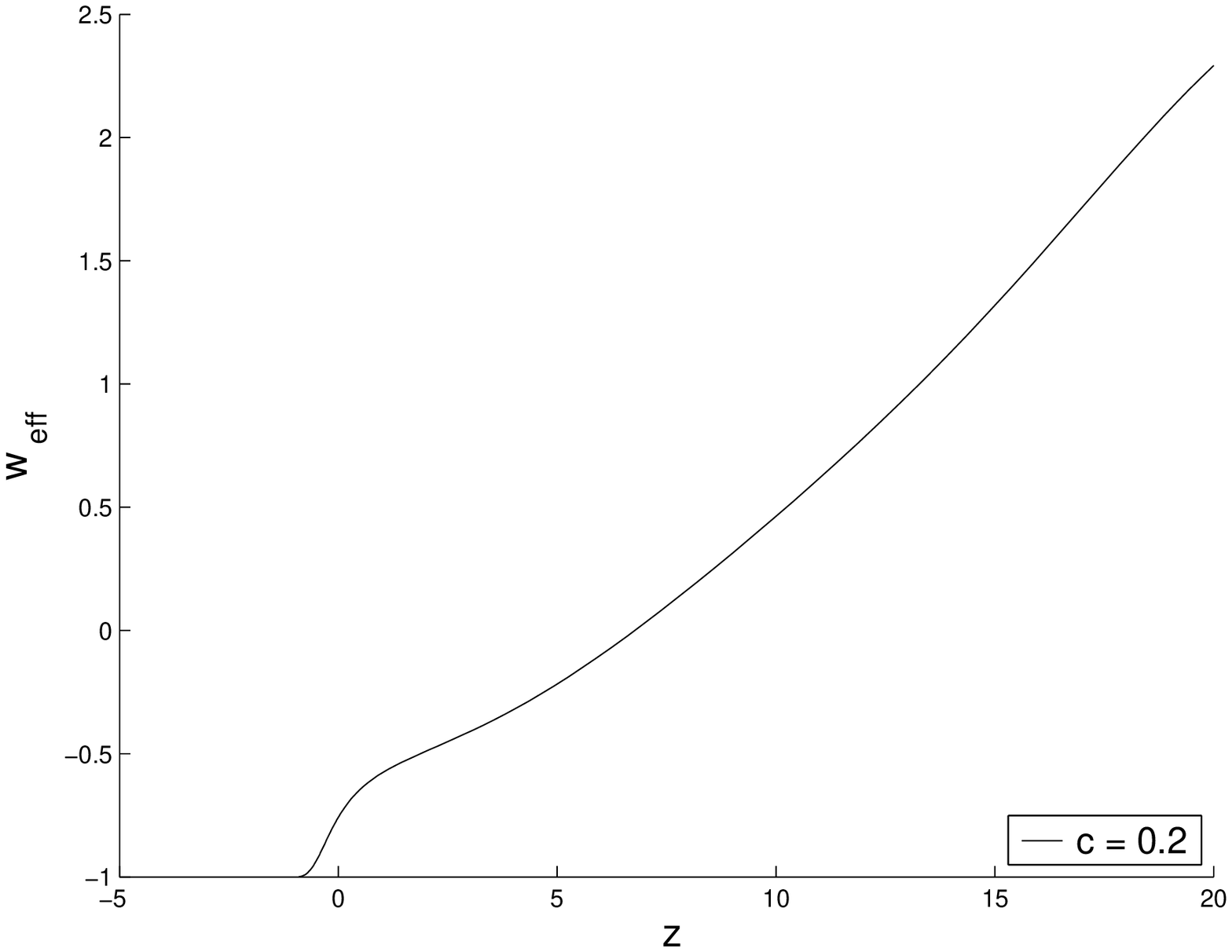}
\includegraphics[scale=0.40]{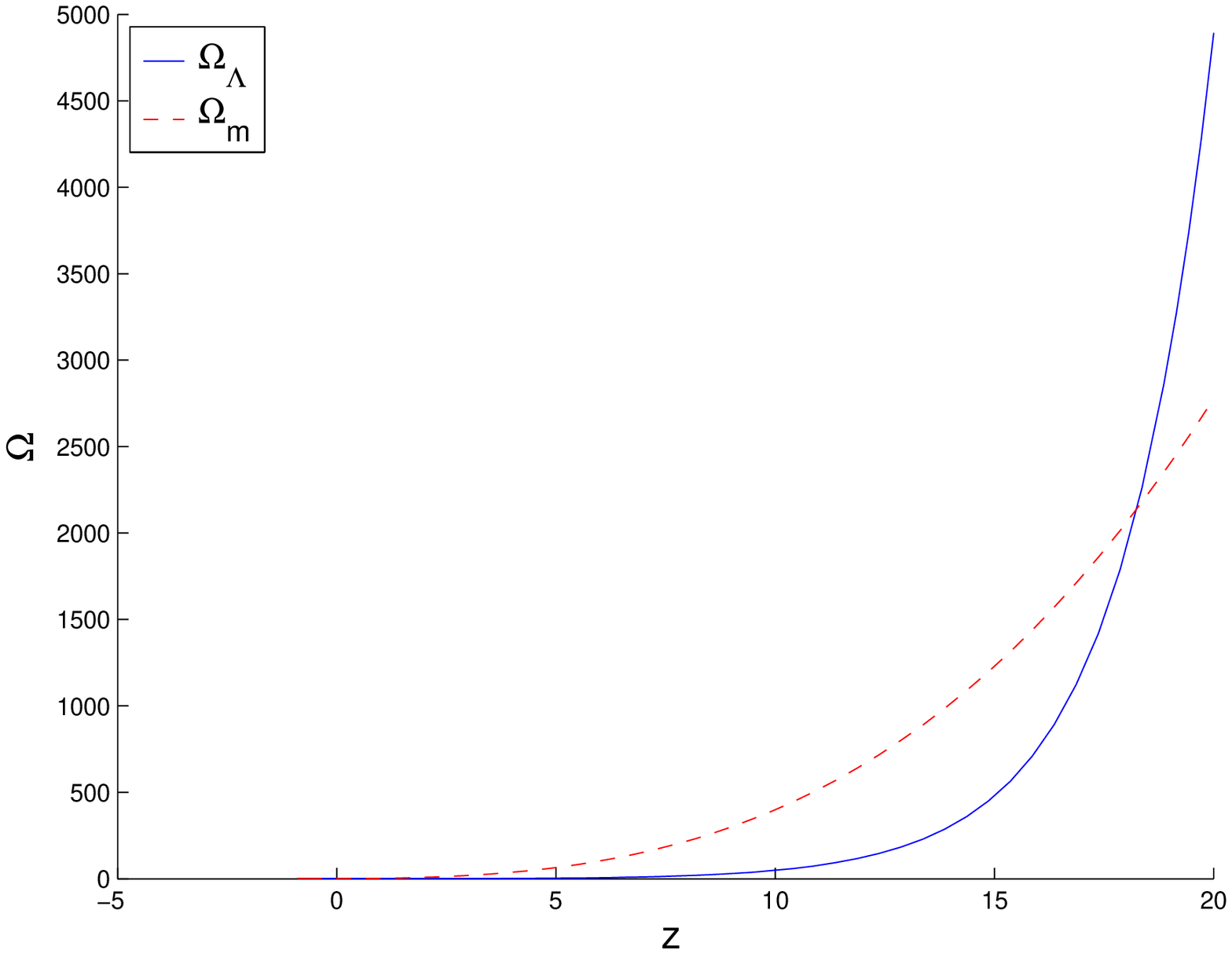}
\caption[]{\small The evolution of $w_{\rm eff}$.  $L=R_{\rm ph}$,
$\epsilon=+1$, $\Omega_m=0.3$ and $\Omega_{r_c}=0.12$. For $c=0.2$
or smaller, $w_{\rm eff}$ grows larger and becomes positive (upper)
at high redshift region. This implies that the vacuum energy
dominates over matter as z grows larger (lower), which would violate
the success of the standard Big Bang cosmology and therefore is not
realistic. } \label{fig:w_eff_ph_1a}
\end{center}
\end{figure}

\begin{figure}
\begin{center}
\includegraphics[scale=0.40]{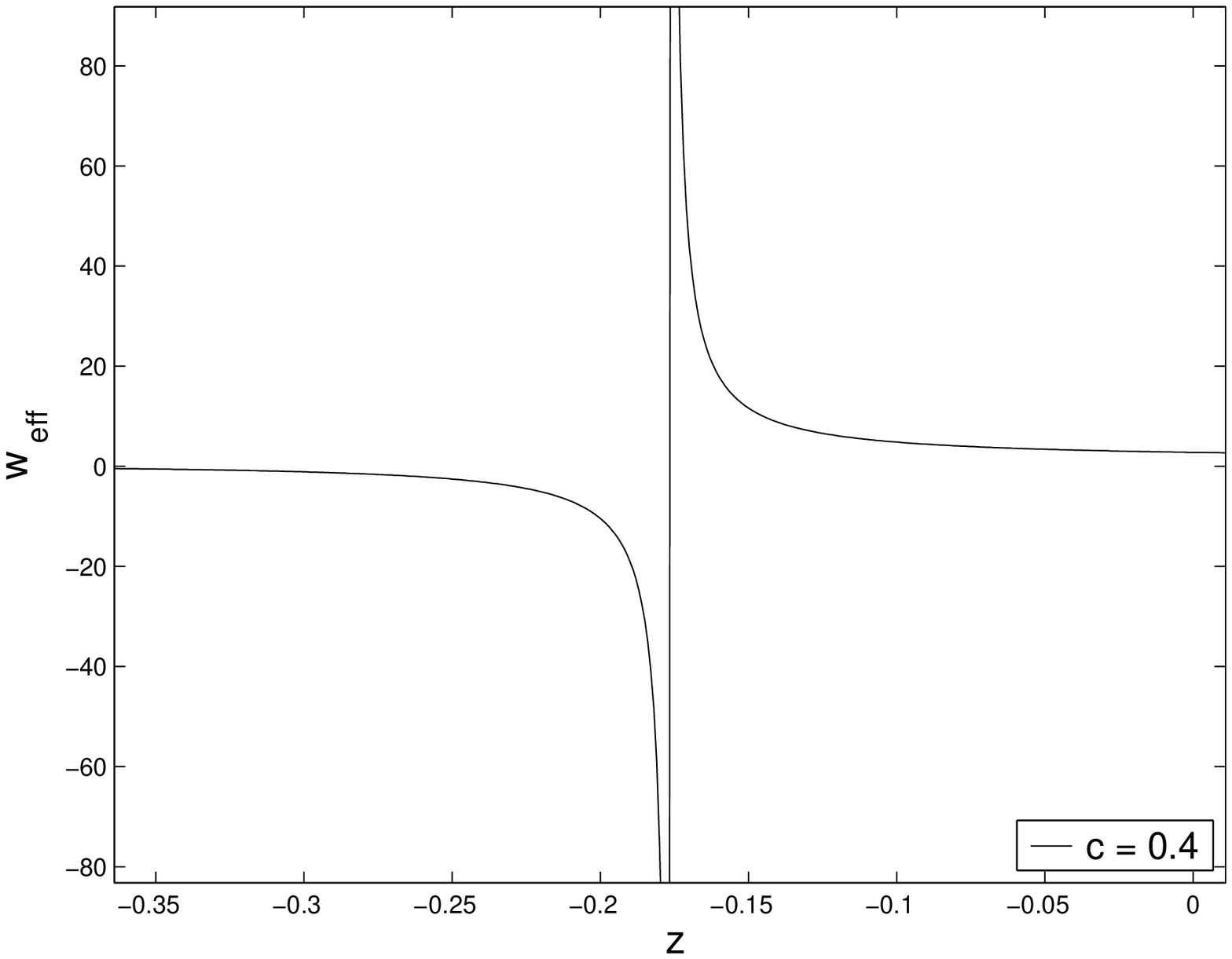}
\caption[]{\small The evolution of $w_{\rm eff}$.  $L=R_{\rm ph}$.
$\Omega_m=0.3$, $\Omega_{r_c}=0.12$ and $\epsilon=-1$. As the case
shown in FIG.\ref{fig:w_eff_H_2}, there exists a pole in the future
evolution of the EoS, which implies the breakdown of the effective
description.} \label{fig:w_eff_ph_2}
\end{center}
\end{figure}

\begin{figure}
\begin{center}
\includegraphics[scale=0.40]{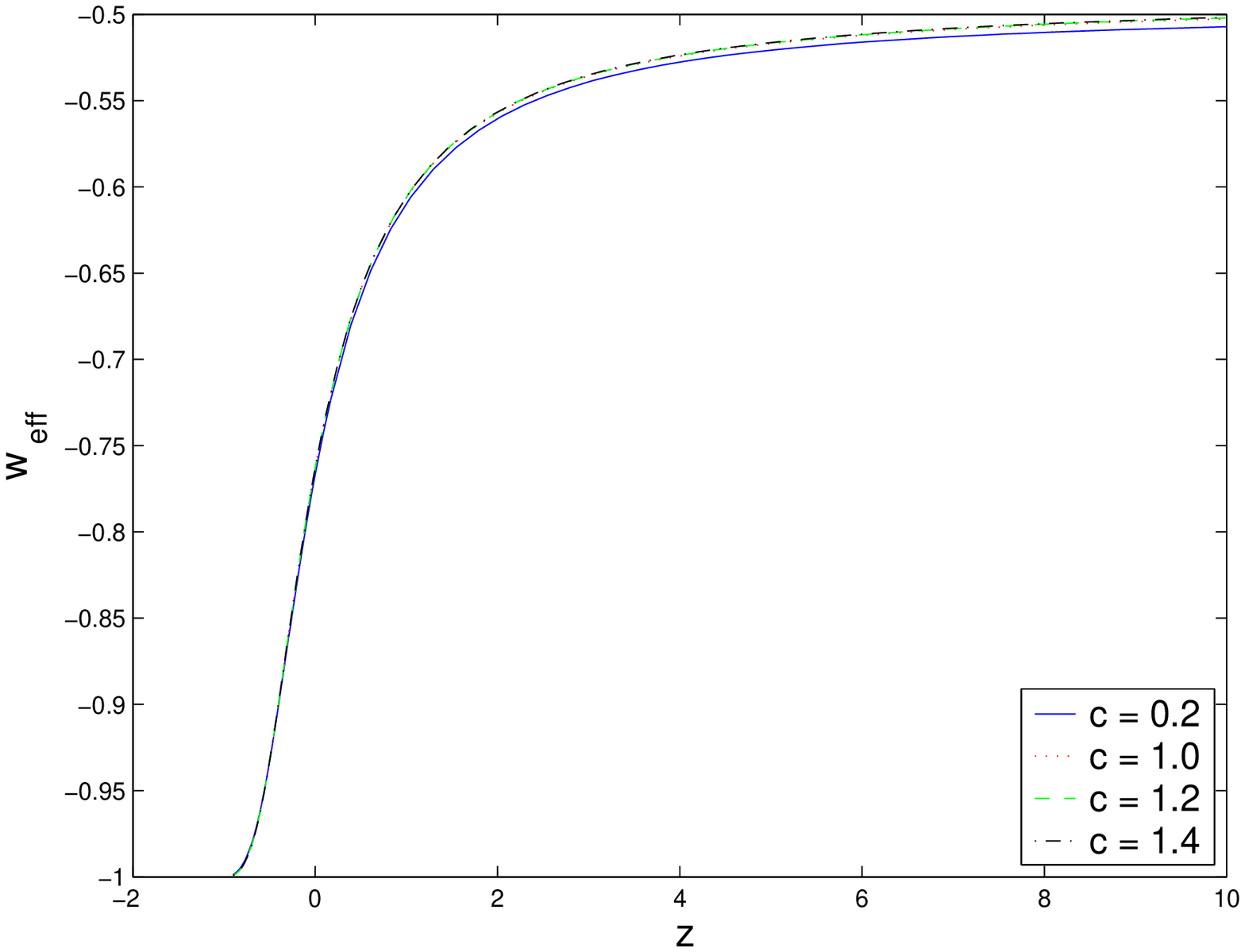}
\caption[]{\small The evolution of $w_{\rm eff}$.  $L=R_{\rm eh}$.
$\Omega_m=0.3$, $\Omega_{r_c}=0.12$ and $\epsilon=+1$}
\label{fig:w_eff_eh_1}
\end{center}
\end{figure}

\begin{figure}
\begin{center}
\includegraphics[scale=0.40]{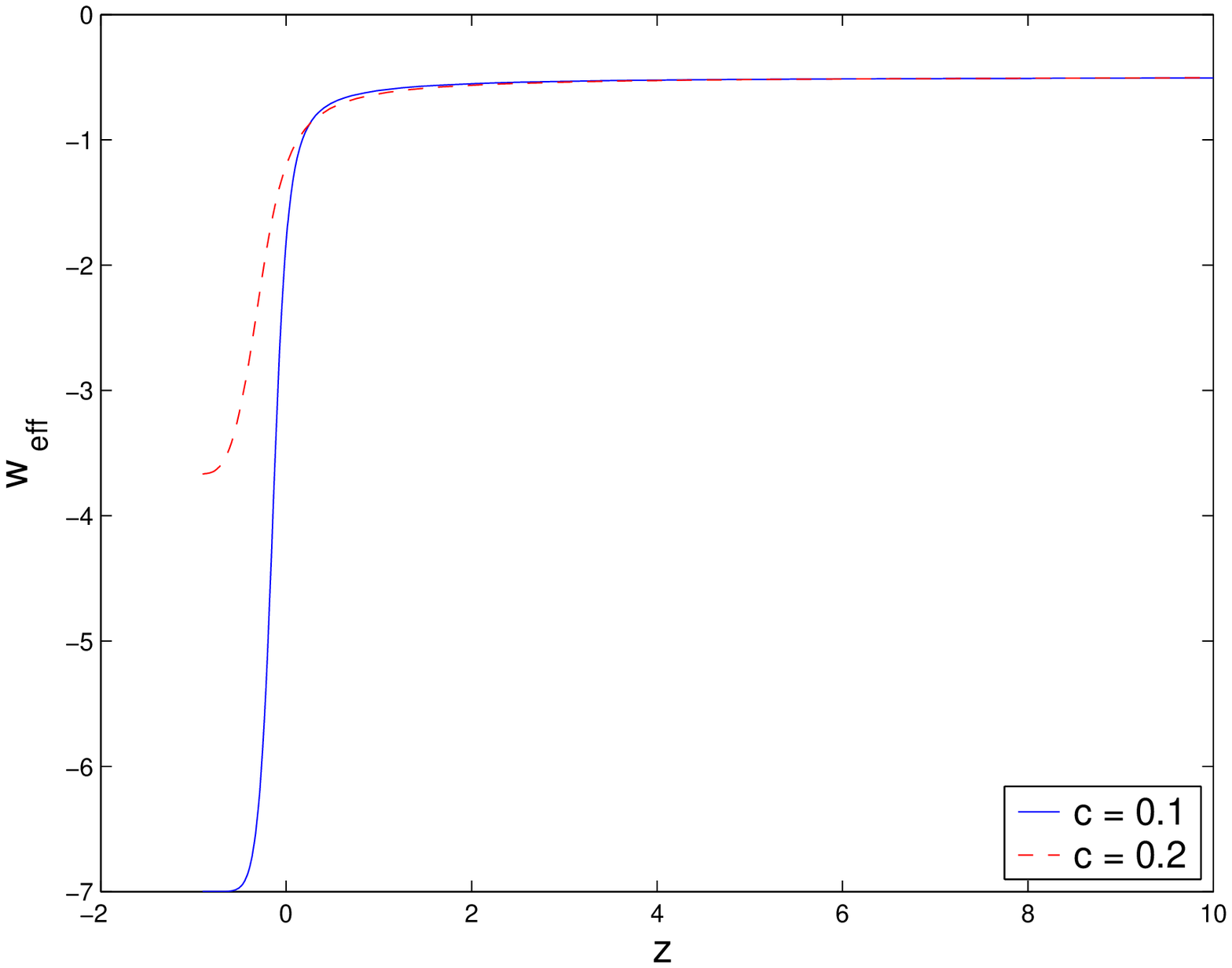}
\includegraphics[scale=0.40]{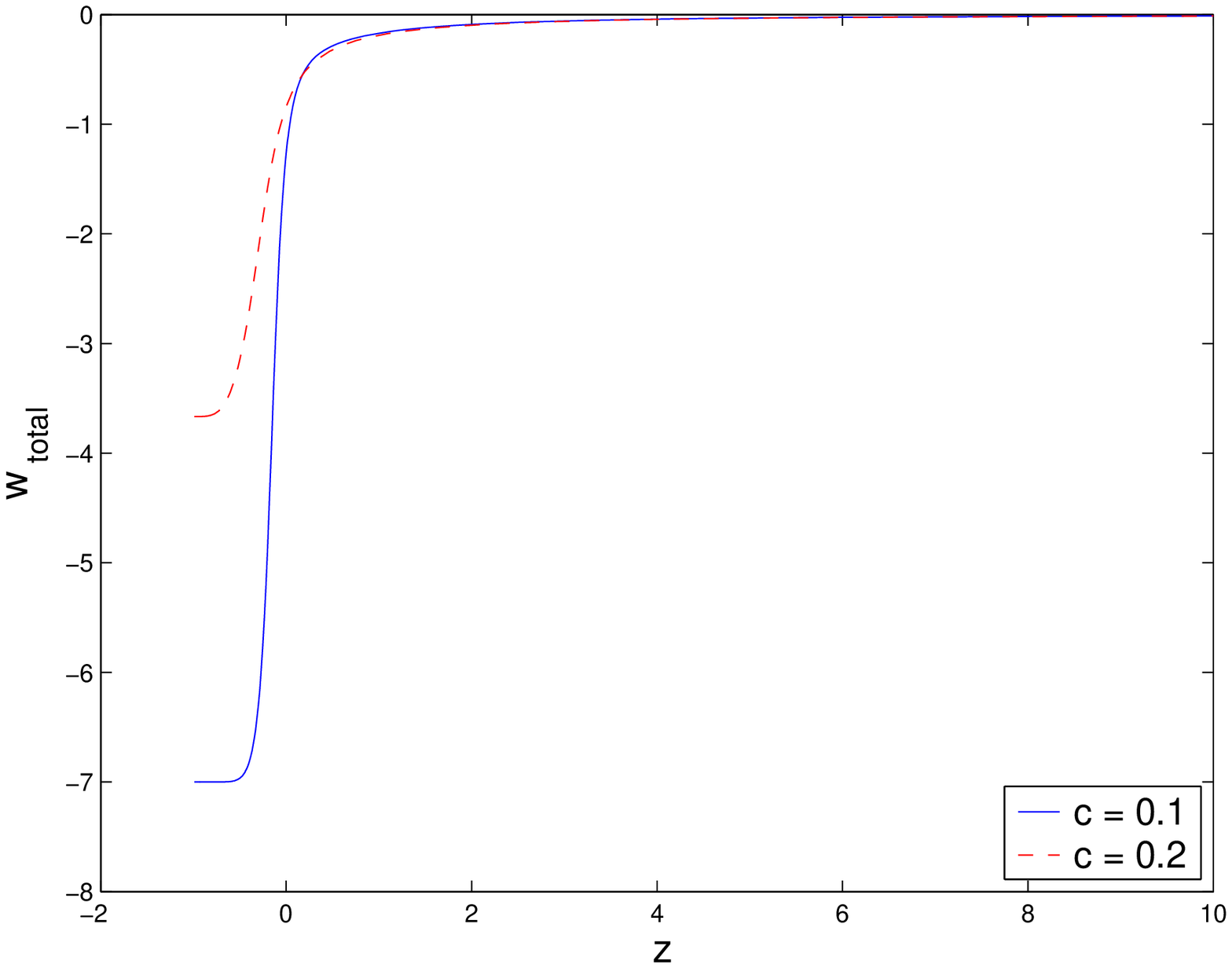}
\caption[]{\small The evolution of $w_{\rm eff}$.  $L=R_{\rm eh}$,
$\epsilon=+1$, $\Omega_m=0.3$ and $\Omega_{r_c}=0.06$.  }
\label{fig:w_eff_eh_1a}
\end{center}
\end{figure}

\begin{figure}
\begin{center}
\includegraphics[scale=0.40]{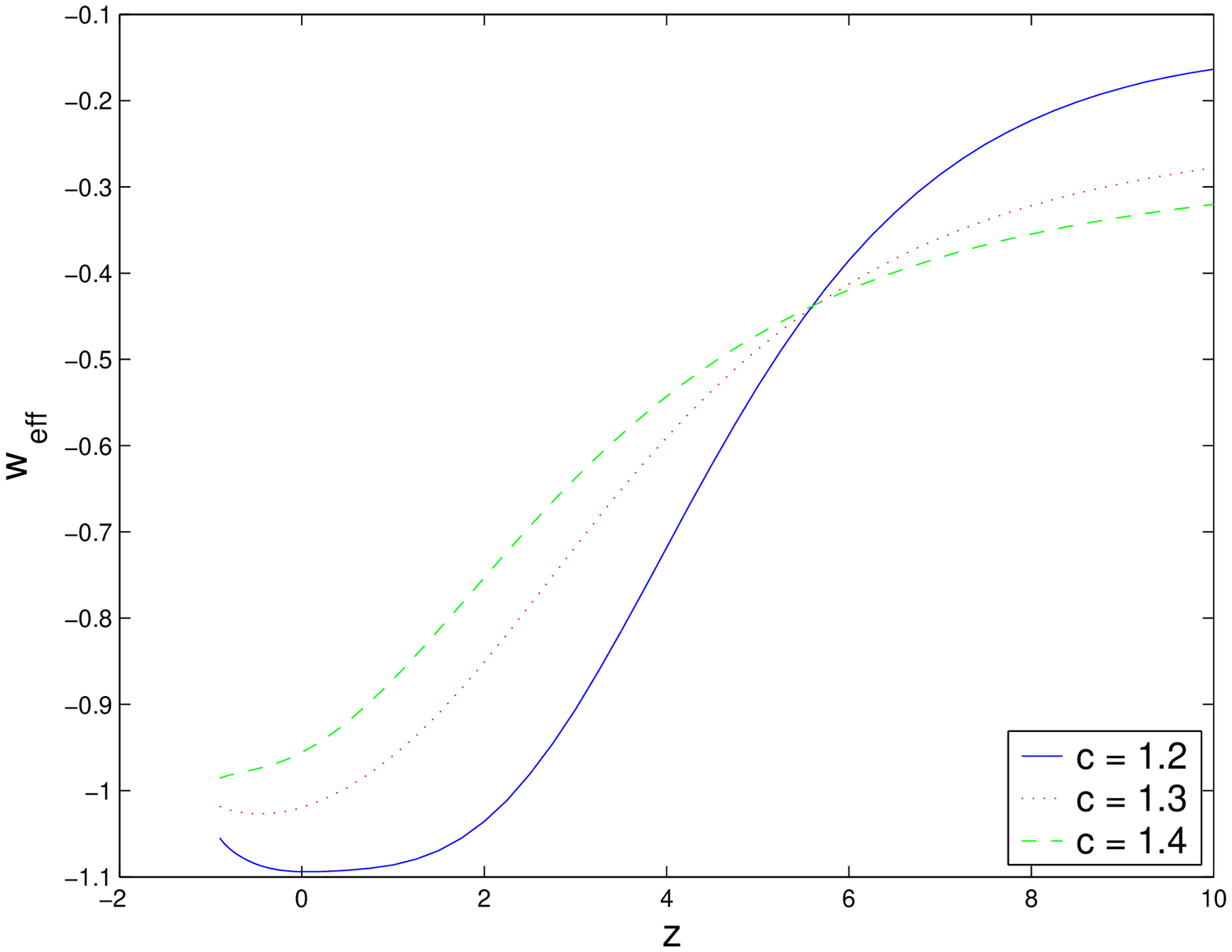}
\includegraphics[scale=0.40]{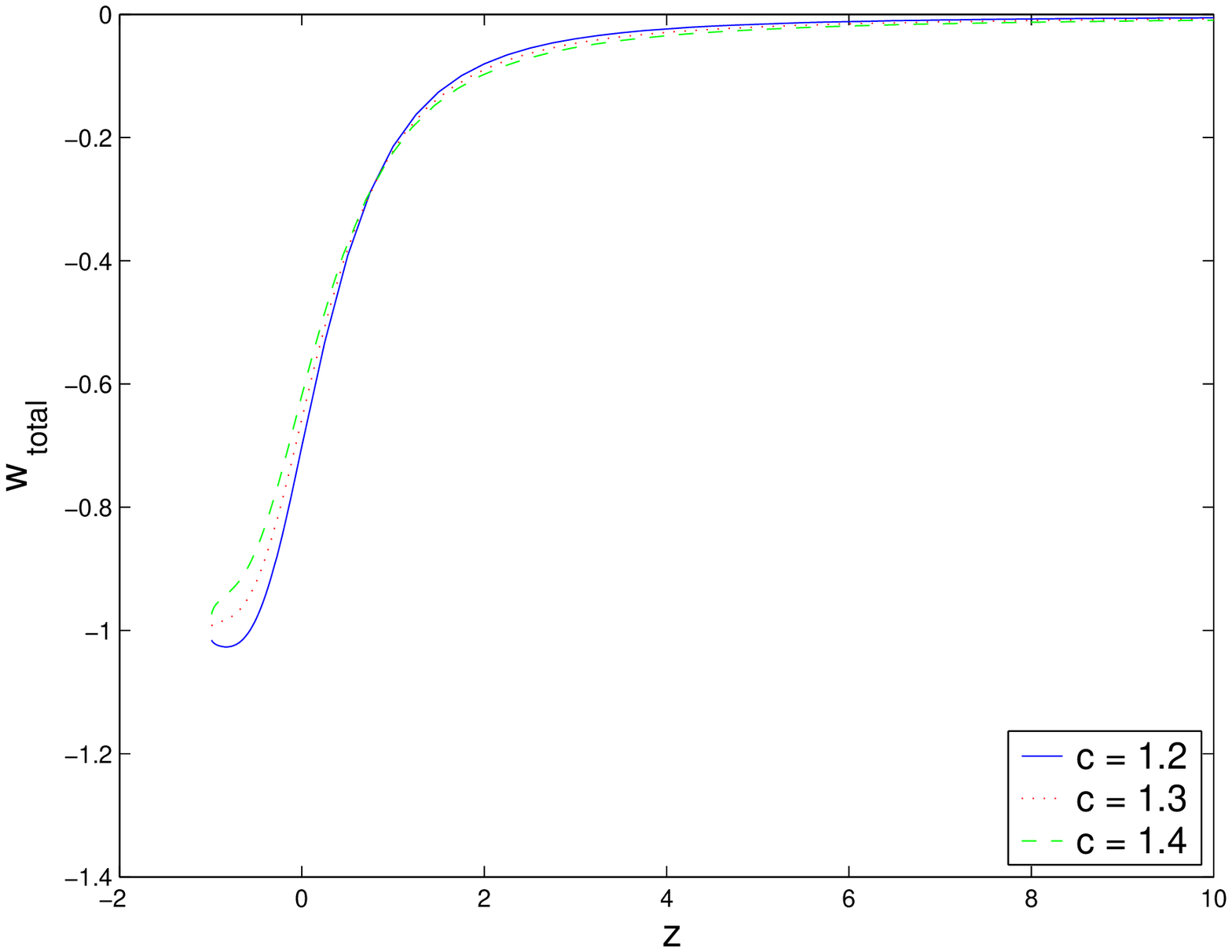}
\caption[]{\small The evolution of $w_{\rm eff}$.  $L=R_{\rm eh}$,
$\epsilon=-1$, $\Omega_m=0.3$ and $\Omega_{r_c}=0.12$. The effective
EoS crosses $-1$, and since it dominates over matter in the future,
 the total EoS also crosses $-1$. This means that the future
Big Rip singularity is not avoidable in this case. }
\label{fig:w_eff_eh_2}
\end{center}
\end{figure}

\begin{figure}
\begin{center}
\includegraphics[scale=0.40]{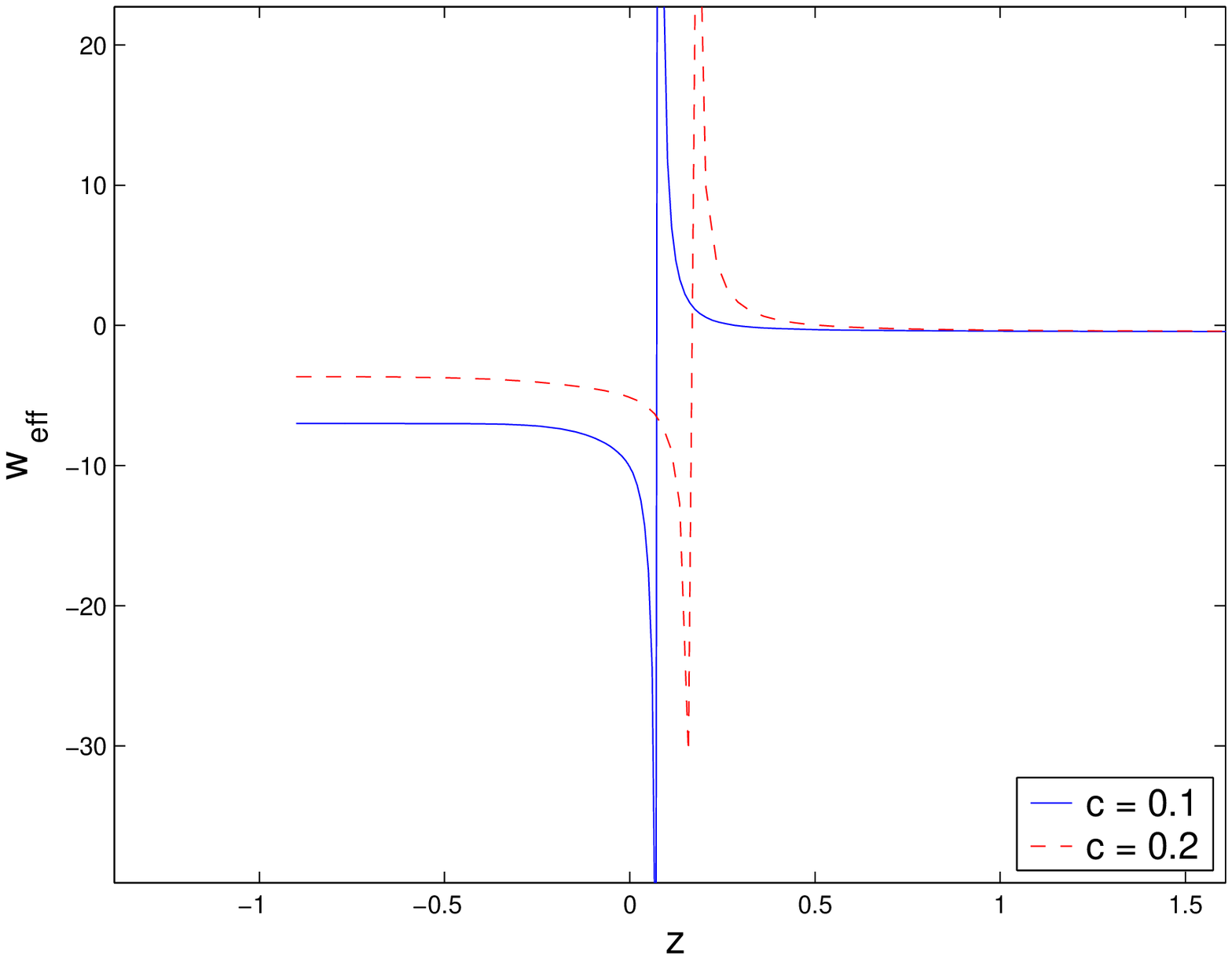}
\includegraphics[scale=0.40]{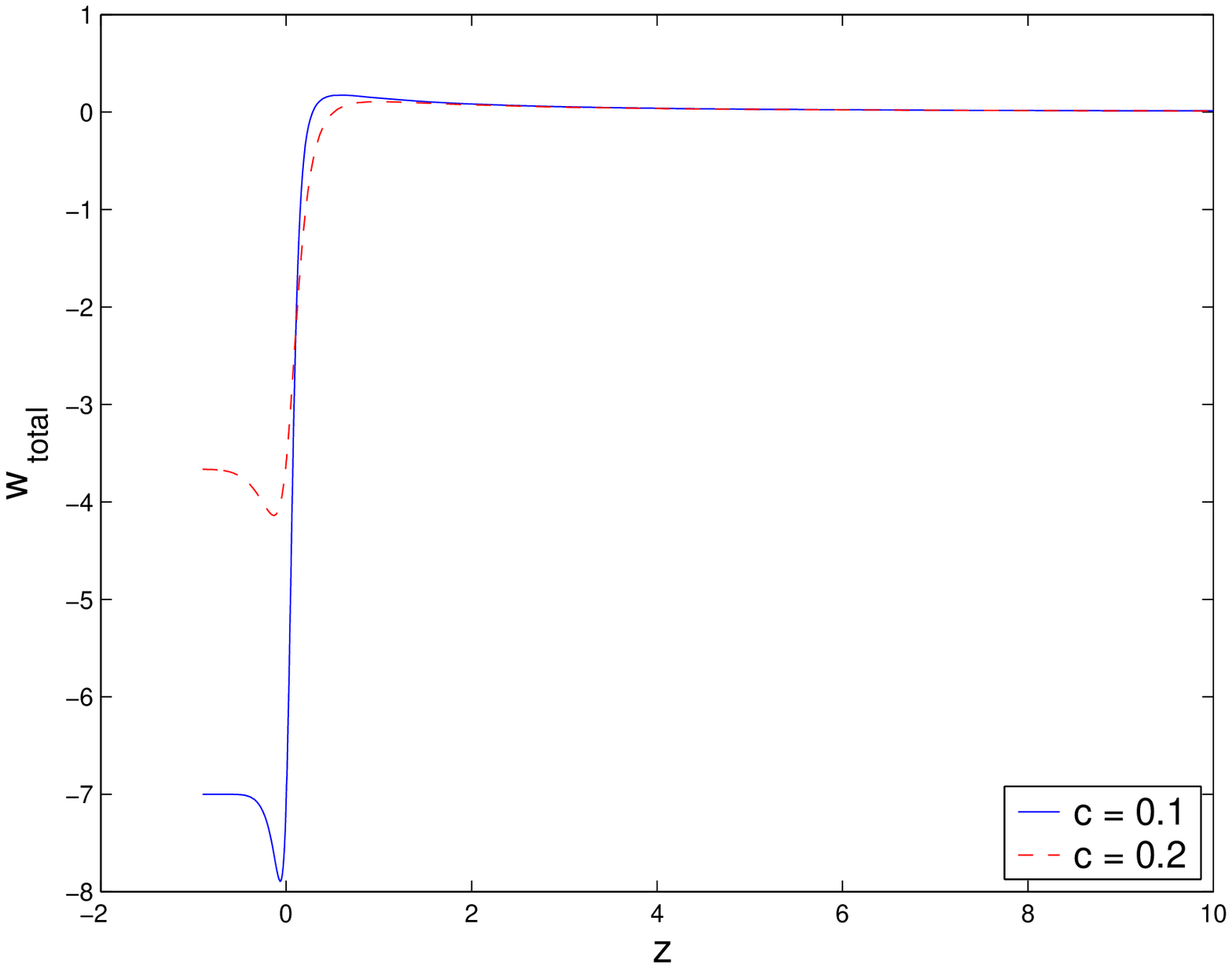}
\caption[]{\small The evolution of $w_{\rm eff}$.  $L=R_{\rm eh}$,
$\epsilon=-1$, $\Omega_m=0.3$ and $\Omega_{r_c}=0.06$.  }
\label{fig:w_eff_eh_2a}
\end{center}
\end{figure}

\section{Parameter fitting with Type Ia Supernova observation and Baryon Acoustic Oscillations }\label{sec3}
In this section, we  confront our model with observational data and
constrain the parameters. We use the SNe data compiled by Davis et
al.\cite{Davis}, which consists of 192 SNe classified as SNIa with
redshift up to $z=1.755$. This dataset is a combination of several
subsets which are 45 nearby SNe\cite{nearby SNe}, 60 SNe from
ESSENCE\cite{ESSENCE}, 57 SNe from the Supernova Legacy Survey
(SNLS)\cite{SNLS} and 30 high redshift SNe from Hubble Space
Telescope (HST)\cite{HST}, of which the data from SNLS and the
nearby SNe were refitted in \cite{ESSENCE}. What the supernova
observations provide is the distance modulus $\mu_{obj}$. This
quantity can be calculated from the model by \eq
\mu_{th}(z;\{\theta_k\})=5
\textrm{lg}D_L(z;\{\theta_k\})+\mathcal{M} ,\ee where $\{\theta_k\}$
represents the parameters of the model: $\{\Omega_{m0},c\}$ for
$L=H^{-1}$, and $\{\Omega_{m0},\Omega_{r_c},c\}$ for $L=R_{\rm ph}\
\textrm{or} \ R_{\rm eh}$. $\mathcal{M}$ is a nuisance parameter
consisting of the Hubble constant $H_0$ and the absolute magnitude
$M$ \eq \mathcal{M}=M+5\textrm{lg}(\frac{c/H_0}{1Mpc})+25 \ee (here
c denotes the speed of light). $D_L$ is the dimensionless luminosity
distance free of Hubble constant. In a spatially flat universe it is
defined by \eq D_L=(1+z)\int^z_0\frac{dz'}{E(z')} .\ee The best fits
are obtained by minimizing the quantity \eq
\chi^2(\{\theta_k\},\mathcal{M})=\sum^{192}_i\frac{(\mu_{obs}-\mu_{th}(z_i;\{\theta_k\},\mathcal{M})^2}{\sigma^2_i}
,\ee where $\sigma_i$ are the observational uncertainties. We
actually deal with the quantity $\chi^2$ with the nuisance parameter
marginalized over \eq \hat{\chi}^2=-2\textrm{ln}\int
e^{-\chi^2/2}d\mathcal{M} .\ee This is equivalent to minimize
$\chi^2$ with respect to $\mathcal{M}$\cite{statistics}, up to a
negligible constant. One can easily show that $\chi^2$ can be
expanded in $\mathcal{M}$ around $\mathcal{M}=0$ by  \eq
\chi^2(\{\theta_k\},\mathcal{M})=\chi^2(\mathcal{M}=0,\{\theta_k\})-2B\mathcal{M}+C\mathcal{M}^2
,\ee where \eq
B=\sum^{192}_i\frac{\mu_{obs}(z_i)-\mu_{th}(z_i;\{\theta_k\},\mathcal{M}=0)}{\sigma_i^2}
,\ee \eq C=\sum^{192}_i\frac{1}{\sigma^2_i} .\ee Obviously $\chi^2$
becomes minimized if $\mathcal{M}=B/C$. Therefore, in practice we
use \eq
\chi^2_{SN}(\{\theta_k\})=\chi^2(\mathcal{M}=0,\{\theta_k\})-\frac{B^2}{C}
\ee as an alternative to $\hat{\chi}^2$ for the sake of efficiency
in practical calculation without losing accuracy.

Another observational data we resort to as a complement to SNe data
is from the observations of baryon acoustic oscillation peak(BAO).
The acoustic oscillations in the relativistic plasma at the
recombination epoch may imprint on the power spectrum of the
non-relativistic matter of the late universe. And this acoustic
signature in the large scale clustering of galaxies was detected by
Eisenstein et al.\cite{BAO} using a large sepctroscopic sample of
the luminous red galaxies (LRGs) from the Sloan Digital Sky Survey
(SDSS)\cite{SDSS}. We use the model-independent parameter A as given
in\cite{BAO} \eq
A=\sqrt{\Omega_{m0}}z_1^{-1}[\frac{z_1}{E(z_1)}\int_0^{z_1}\frac{dz}{E{z}}]^{1/3}
,\ee where $z_1=0.35$ is the typical redshift of the LRGs. The
measured value of A is $0.469\pm0.017$\cite{BAO}. Correspondingly
the quantity $\chi^2$ is \eq
\chi^2_{BAO}=\frac{(A-0.469)^2}{0.017^2} .\ee And in the following
we perform the joint analysis using
$\chi^2=\chi^2_{SN}+\chi^2_{BAO}$.

\begin{figure}[htbp]
\begin{center}
\includegraphics[scale=0.40]{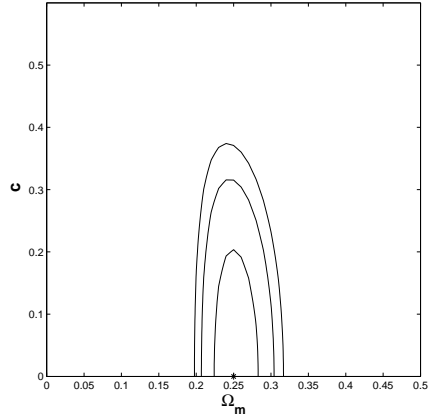}
\caption[]{\small Contour plot within $3\sigma$ from a joint
analysis for $L=H^{-1}$ and $\epsilon=+1$. The best fits are
$\Omega_{m0}=0.25 \pm0.02$ and $c=0+0.14$. } \label{fig:chi2_H_com}
\end{center}
\end{figure}

\begin{figure}[htbp]
\begin{center}
\includegraphics[scale=0.30]{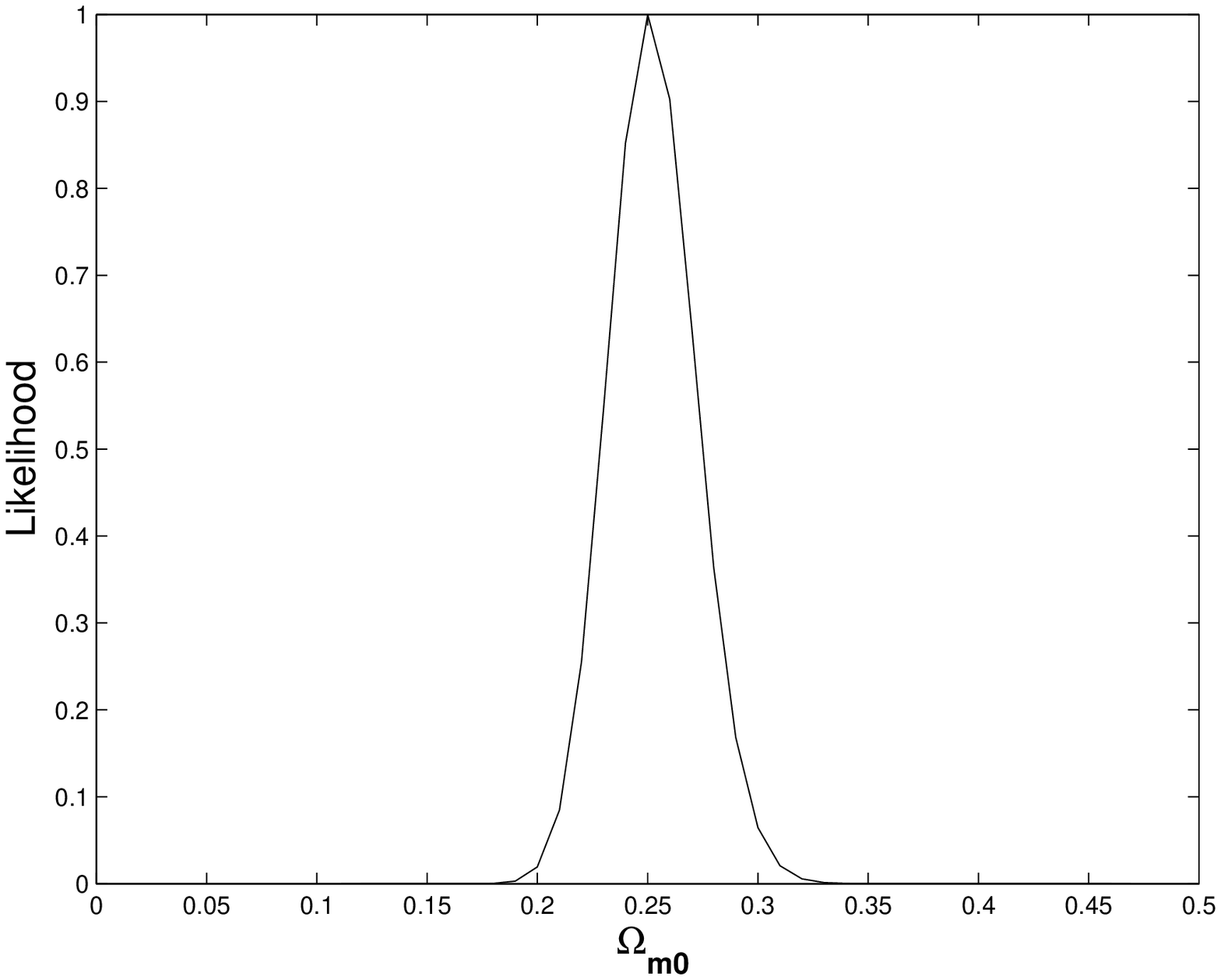}
\includegraphics[scale=0.30]{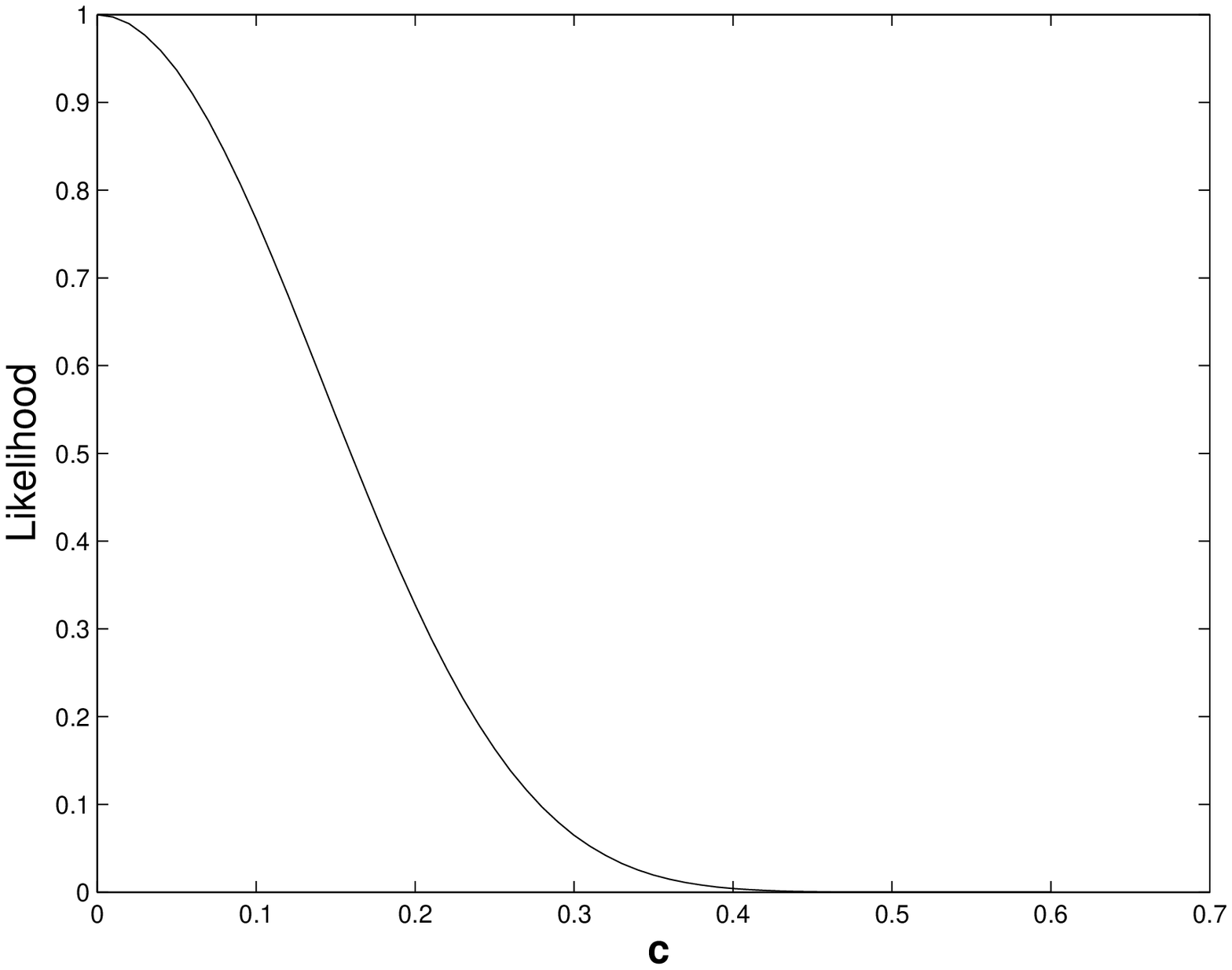}
\caption[]{\small Marginalized likelihood functions for
$\Omega_{m0}$ and $c$. } \label{fig:likelihood_H}
\end{center}
\end{figure}

\begin{figure}[htbp]
\begin{center}
\includegraphics[scale=0.40]{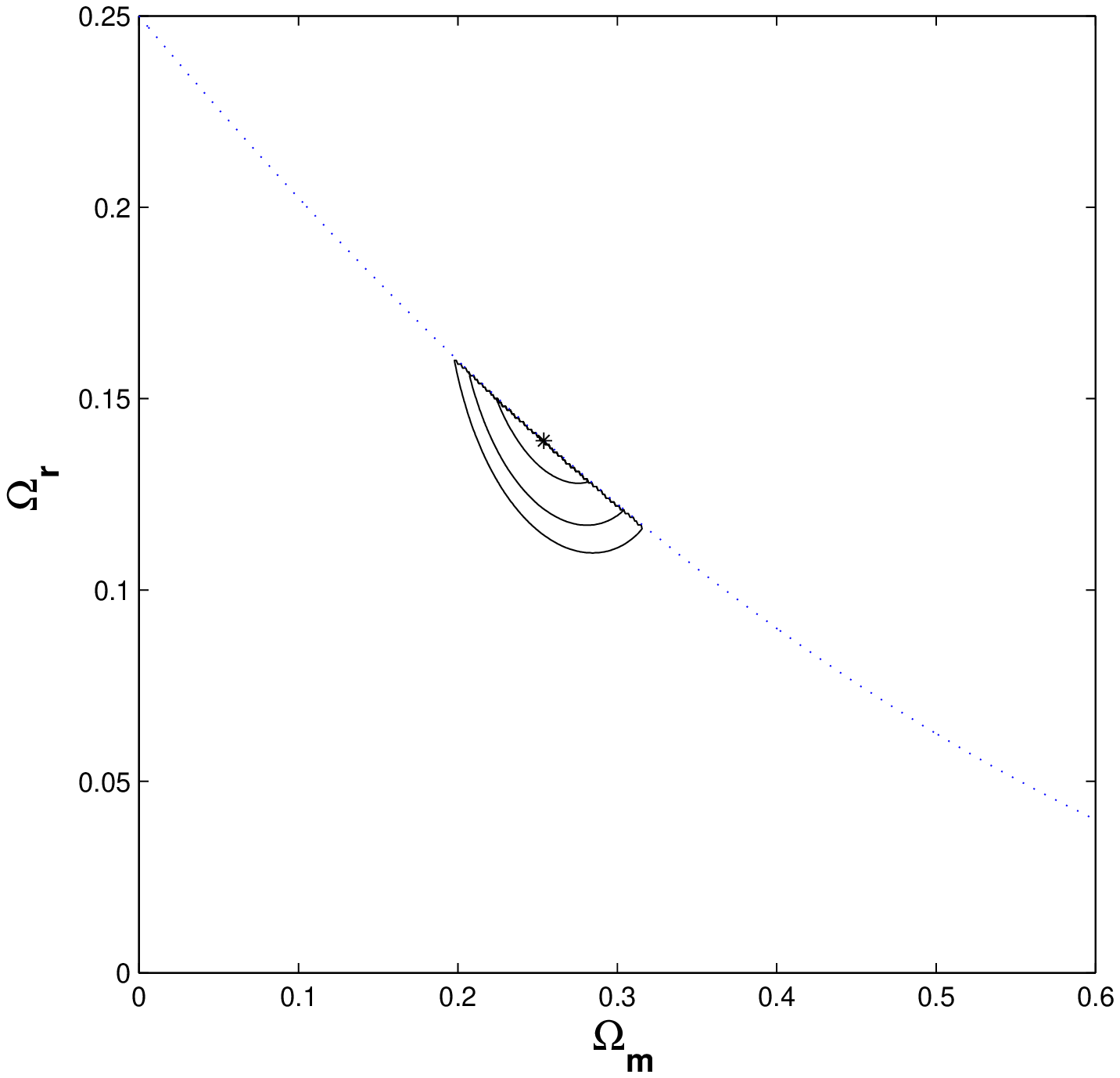}
\caption[]{\small Contour plot within $3\sigma$ from a joint
analysis for $L=R_{\rm ph}$ and $\epsilon=+1$, with c marginalized.
The best fits are $\Omega_{r_c}=0.14\pm 0.01$ and
$\Omega_{m0}=0.25\pm 0.02$. The dotted line denotes the situation of
$\Omega_{\Lambda0}=0$, above which $\Omega_{\Lambda0}<0$ and there
the region is forbidden in our model.} \label{fig:chi2_ph}
\end{center}
\end{figure}

\begin{figure}[htbp]
\begin{center}
\includegraphics[scale=0.30]{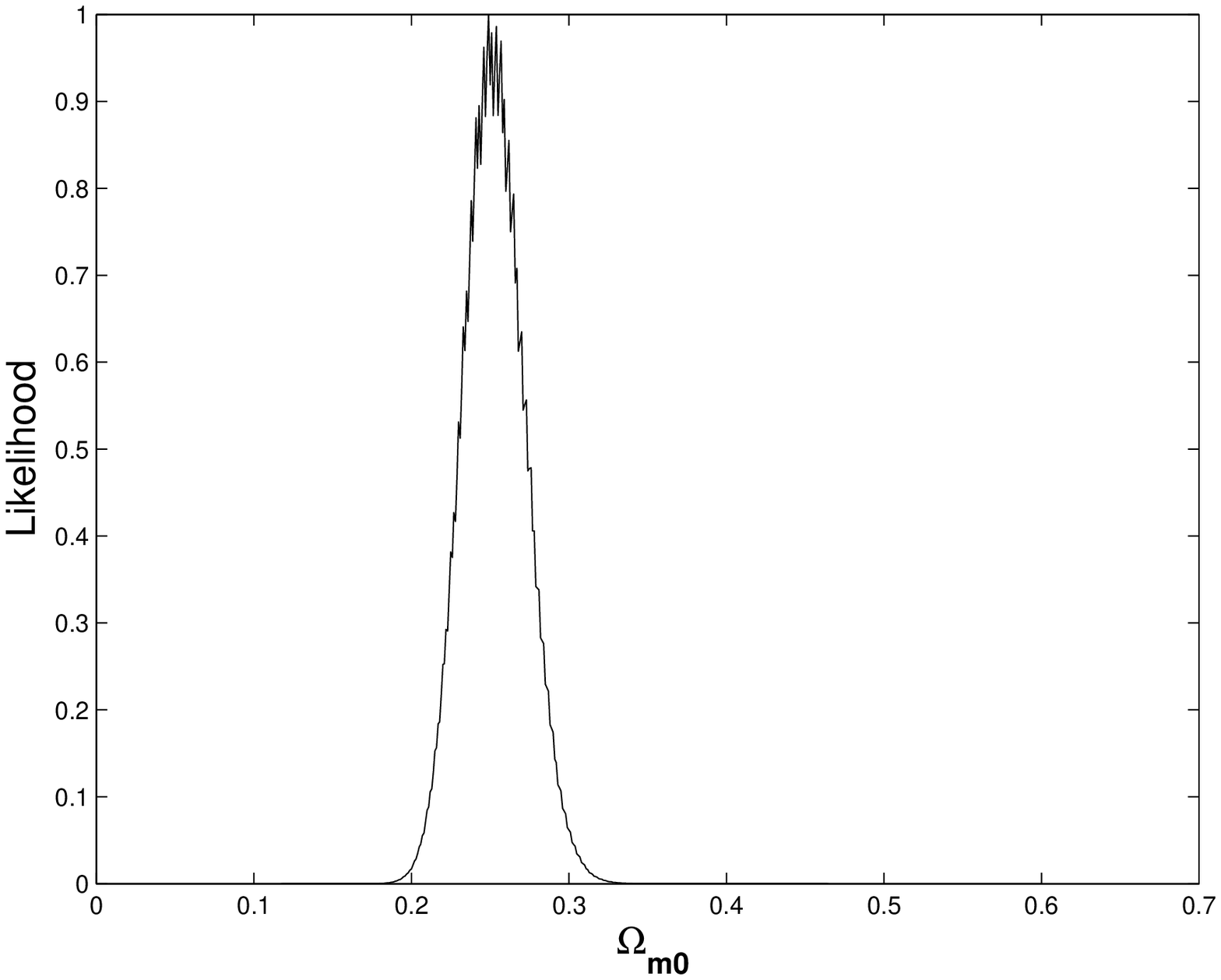}
\includegraphics[scale=0.30]{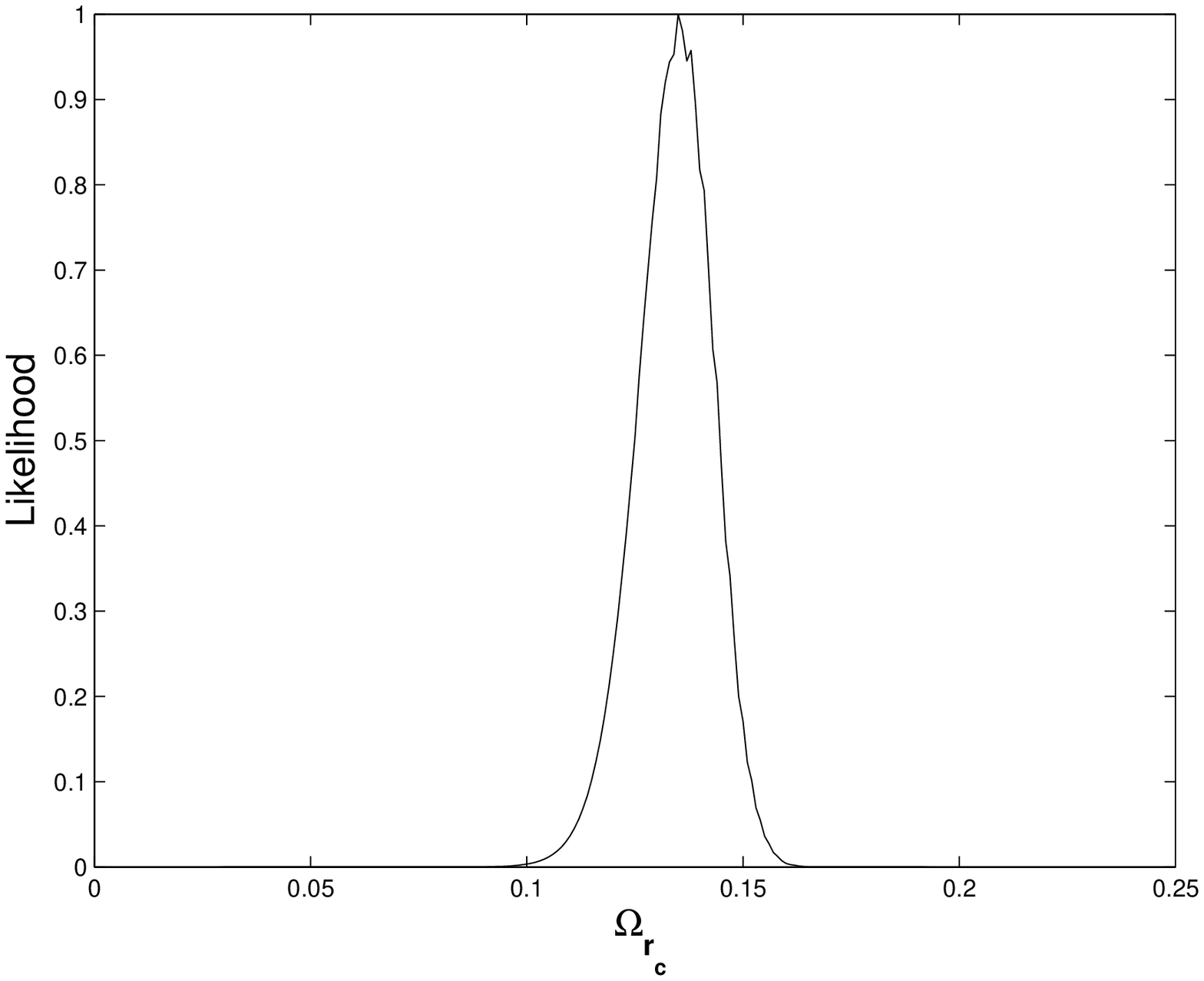}
\caption[]{\small Marginalized likelihood functions for
$\Omega_{m0}$ and $\Omega_{r_c}$. } \label{fig:likelihood_ph}
\end{center}
\end{figure}

\begin{figure}[htbp]
\begin{center}
\includegraphics[scale=0.40]{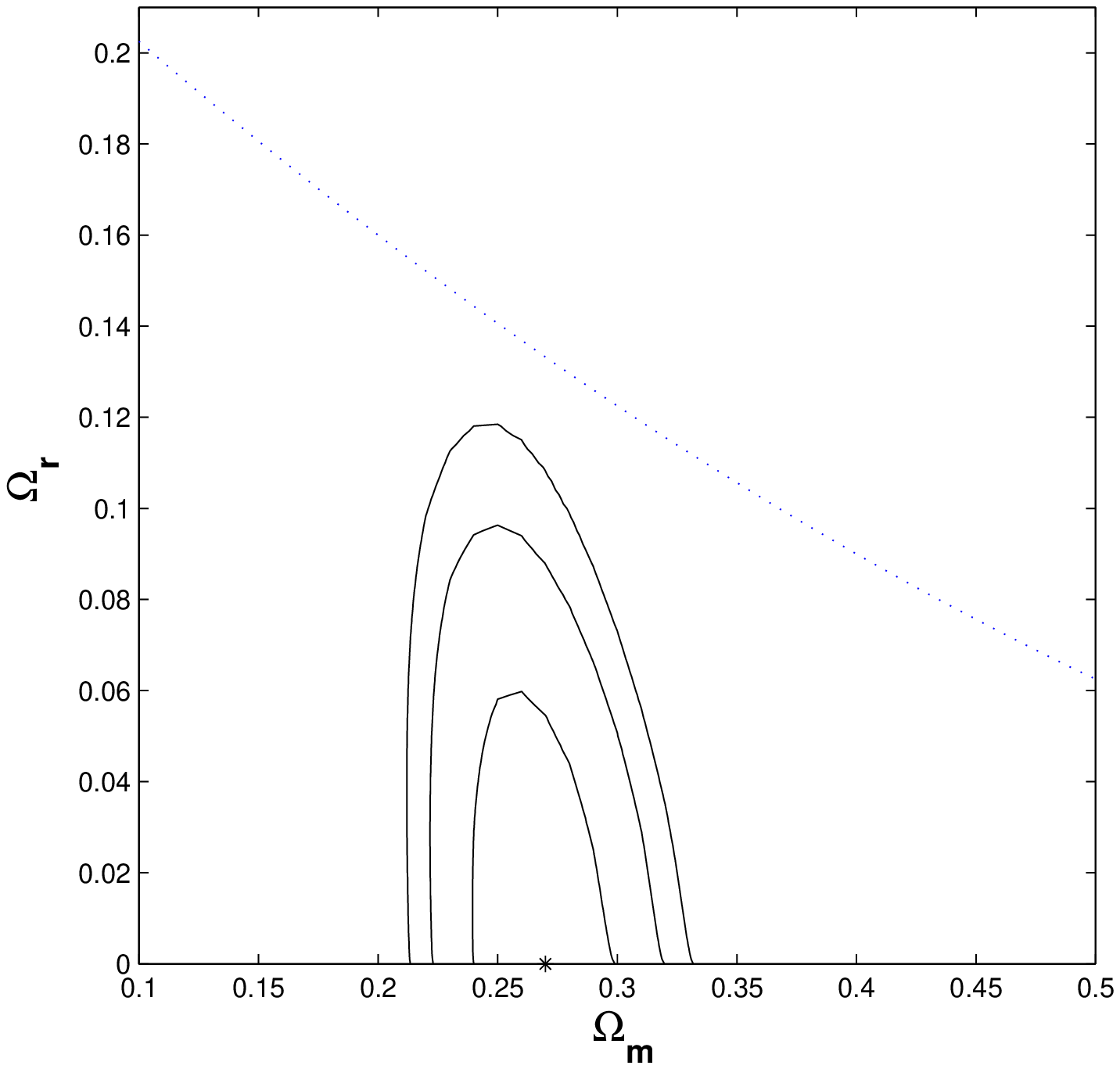}
\caption[]{\small Contour plot within $3\sigma$ from a joint
analysis for $L=R_{\rm eh}$ and $\epsilon=+1$, with c marginalized.
The best fits  are $\Omega_{r_c}=0+0.04$ and
$\Omega_{m0}=0.27\pm0.02$. Corresponding to the best fits, $c=0.76$.
Again, the dotted line denotes the boundary between
$\Omega_{\Lambda0}>0$(below) and $\Omega_{\Lambda0}<0$(above). }
\label{fig:chi2_eh1}
\end{center}
\end{figure}

\begin{figure}[htbp]
\begin{center}
\includegraphics[scale=0.30]{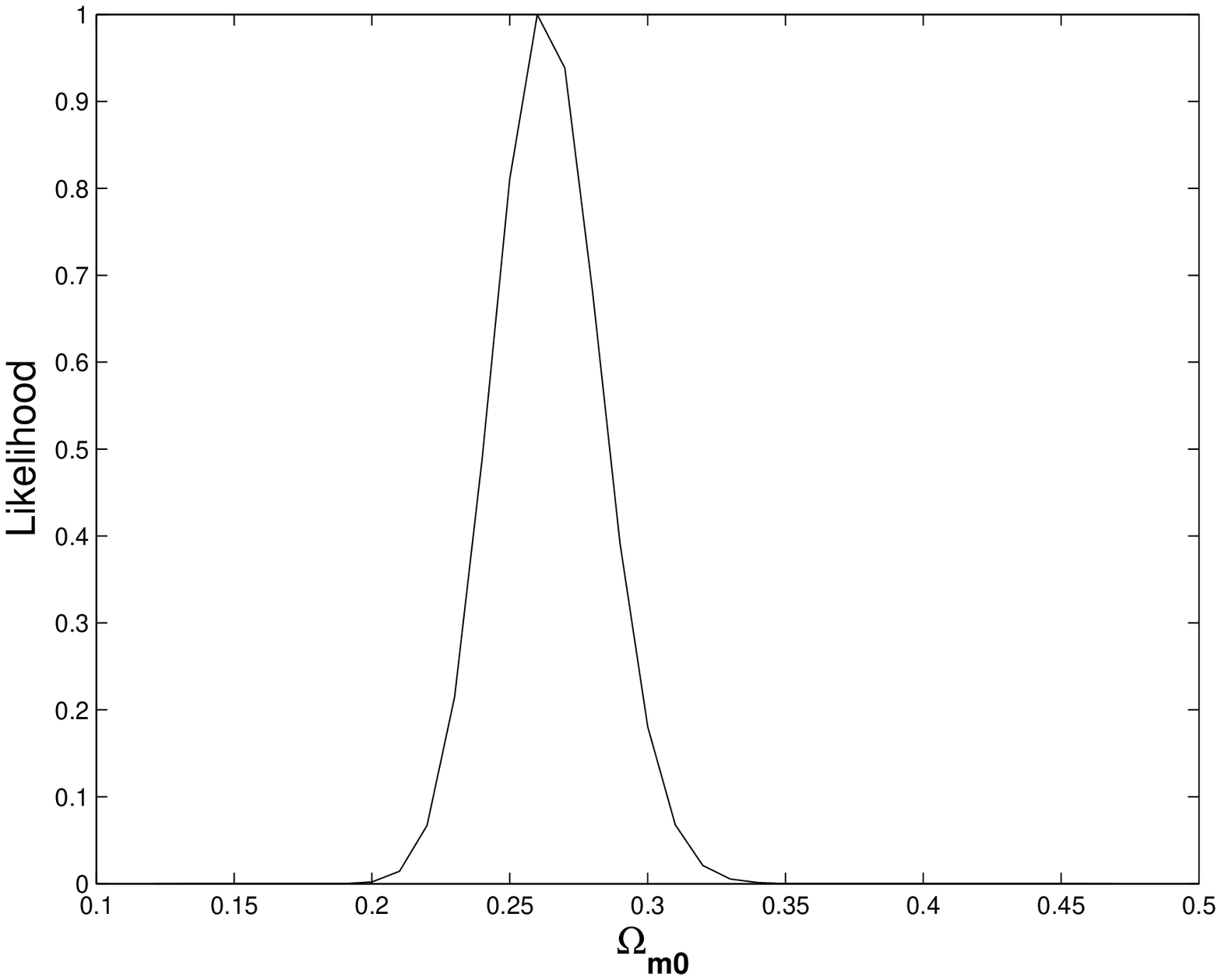}
\includegraphics[scale=0.30]{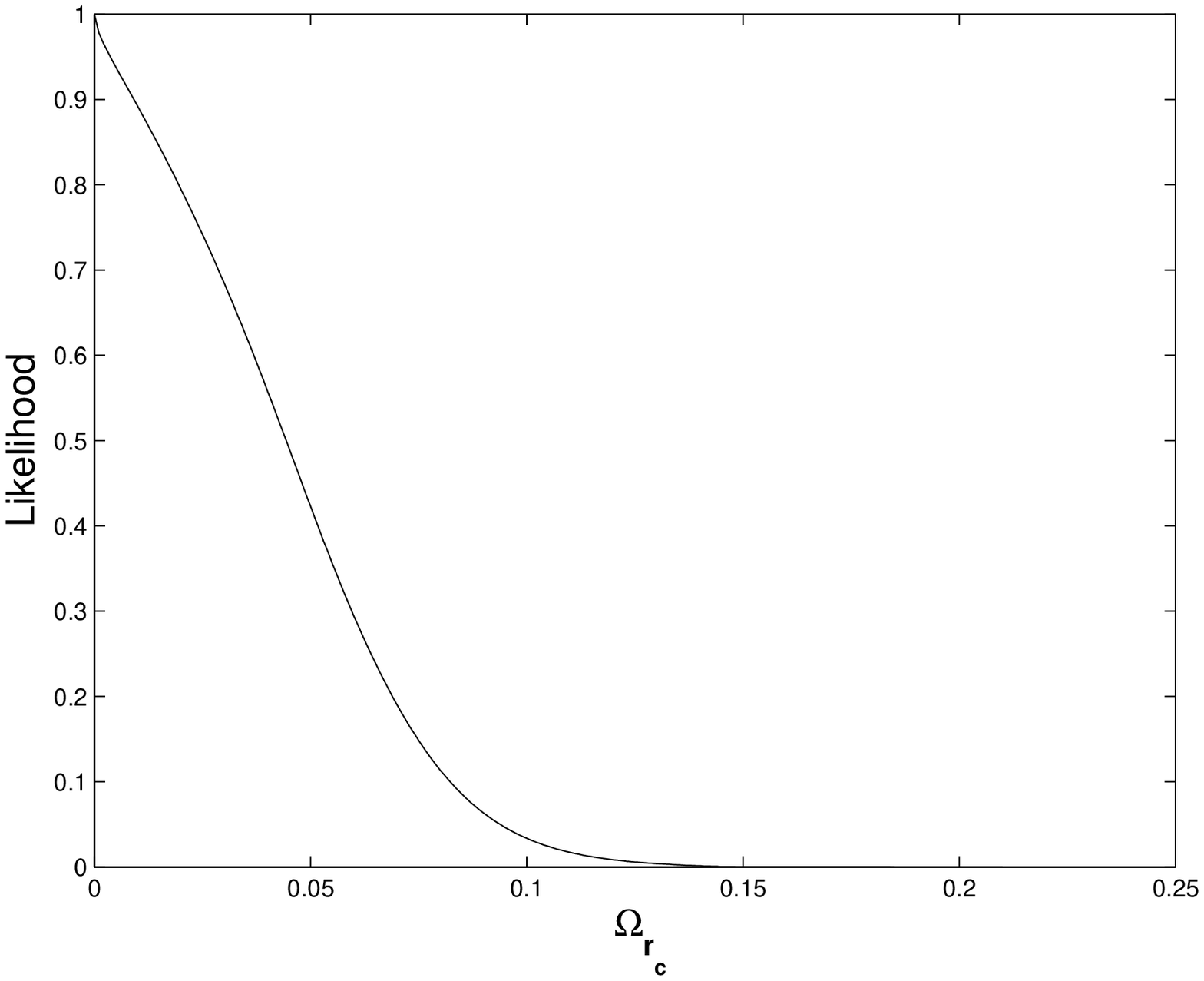}
\caption[]{\small Marginalized likelihood functions for
$\Omega_{m0}$ and $\Omega_{r_c}$.  } \label{fig:likelihood_eh_1}
\end{center}
\end{figure}

\begin{figure}[htbp]
\begin{center}
\includegraphics[scale=0.40]{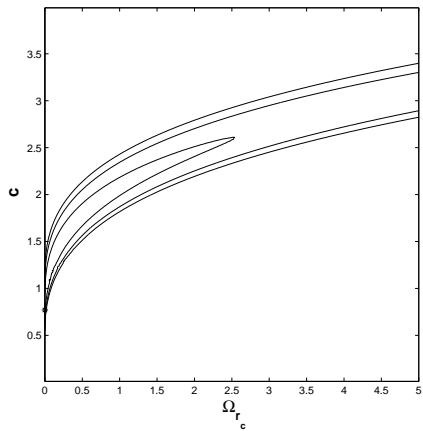}
\caption[]{\small Contour plot for $\Delta\chi^2=2.30,6.17,9.21$
from a joint analysis for $L=R_{\rm eh}$ and $\epsilon=-1$, with
$\Omega_{m0}$ marginalized. The best fits  are $\Omega_{r_c}=0$,
$c=0.77$ denoted by a star on the plot. Corresponding to the best
fits, $\Omega_{m0}=0.27$. } \label{fig:chi2_eh2}
\end{center}
\end{figure}

For $L=H^{-1}$, there are two parameters $(\Omega_{m0},c)$ to be
fitted. And $\Omega_{r_c}$ is determined by Eq.(\ref{initial_H}).
The result is shown in Fig.\ref{fig:chi2_H_com}.  The best fits
$\Omega_{m0}=0.25$, $c=0$ and $\Omega_{r_c}=0.14$ indicate that the
observations prefer a pure DGP model without the holographic vacuum
energy. This best fits are also consistent with those obtained in
literature\cite{DGP test}. Fig.\ref{fig:likelihood_H} shows the
marginalized likelihood function for the two parameters, in which
the curve for $\Omega_{m0}$ is near-Gaussian whereas it is highly
asymmetric for $c$ due to the theoretical cutoff of this parameter.
This leads to $\Omega_{m0}=0.25\pm0.02$ and $c<0.14$ within $68.3\%$
confidence level.

For $L=R_{\rm ph}$, the results are given by Fig.\ref{fig:chi2_ph}
and Fig.\ref{fig:likelihood_ph}, where we have marginalized the
parameter $c$. In the contour plot, the dotted line represents
$\Omega_{\Lambda 0}=0$ in Eq.(\ref{initial}). Below this line
$\Omega_{\Lambda 0}>0$. In the region above $\Omega_{\Lambda 0}<0$
and therefore it is the unphysical region for the parameters of the
model. In the pure DGP model, the counterpart of Eq.(\ref{initial})
is \eq\Omega_k=1-2\sqrt{\Omega_{r_c}}-\Omega_{m0}\,
\label{counterpart} ,\ee where $\Omega_k$ denotes the spatial
curvature. If we identify $\Omega_{\Lambda 0}$ in Eq.(\ref{initial})
with $\Omega_k$ in Eq.(\ref{counterpart}), we find that our model in
this case is equivalent to the pure DGP model confined to a
non-closed universe. The best fits are $\Omega_{r_c}=0.14\pm 0.01$
and $\Omega_{m0}=0.25\pm 0.02$, indicating a pure DGP model in a
flat universe without vacuum energy.

For $L=R_{\rm eh}$, we have to consider the two branches. In the (+)
branch, the results are shown in Fig.\ref{fig:chi2_eh1} and
Fig.\ref{fig:likelihood_eh_1}, with the parameter $c$ marginalized.
The best fits are $\Omega_{r_c}=0+0.04$ and
$\Omega_{m0}=0.27\pm0.02$, correspondingly $c=0.76$, indicating a
pure holographic dark energy model with the negligible effect of
higher dimensional gravity. In the (-) branch, as we can see from
Fig.\ref{fig:chi2_eh2} (with $\Omega_{m0}$ marginalized), the two
outmost contours are not closed within a large region of the
parameter space, indicating that current observations cannot impose
tight constraint on the parameters in this case. Therefore the
contours just represent the difference with respect to the minimum
of $\chi^2$, $\Delta\chi^2=2.30,6.17,9.21$ respectively, without an
exact statistical meaning. Thus we do not present the likelihood
plot as before. Despite of this, we can still get the best fits as
$\Omega_{r_c}=0$ and $c=0.77$ with $\Omega_{m0}=0.27$
correspondingly. This also indicates a pure holographic dark energy
model.

\begin{table}
\begin{center}
\caption{Best fits}

\begin{tabular}{lcclll}

L     & $\epsilon$       & $\Omega_{\rm {m0}}$    & $\Omega_{\rm
r_c}$ & $c$     &$w_{{\rm eff}0}$  \\
\hline
$H^{-1}$       & $+1$      &            0.25    &0.14      & 0           &-0.8000\\
$R_{\rm ph}$   & $+1$      &            0.25    &0.14      & N/A         &-0.7015     \\
$R_{\rm eh}$   & $+1$      &            0.27    &0         &0.76          &-1.0828    \\
$R_{\rm eh}$   & $-1$      &            0.27    &0         &0.77         &-1.0731   \\

\end{tabular}
\end{center}
\end{table}

\section{Conclusion}
In this paper, we considered the evolution of the vacuum energy in
the universe described by the DGP model. By numerically studying the
EoS of the vacuum energy and the effective EoS of the combined
effect of both vacuum energy and brane effect, we found that
choosing the IR cut-off as the event horizon, the vacuum energy can
drive the cosmic acceleration in both branches. In addition, the
choice of the Hubble scale as the cut-off can also lead to the
vacuum energy playing the role of dark energy. This is different
from the case in ordinary 4D gravity, where $w_{\Lambda}<-1/3$ only
when the event horizon is chosen as the IR cut-off. Further
investigation shows that when $L=R_{\rm eh}$, the EoS may cross $-1$
and the vacuum energy would end up with a phantom phase, therefore
the Big Rip singularity is inevitable, in contrast to the models
such as LDGP\cite{LDGP} and SDGP\cite{SDGP} where only the effective
EoS posses the crossing behavior and the total EoS is always larger
than $-1$.

Through a joint analysis of SNe data and BAO data, the results of
parameter fitting show that the DGP model with holographic vacuum
energy can be consistent with the joint data constraints within
$68.3\%$ confidence level. For IR cut-off L as the Hubble scale and
the particle horizon in the (+) branch, the best fits indicate that
the observational data prefer a pure DGP model with negligible
vacuum energy. For L as the event horizon in both branches, on the
other hand, the best fits show a preference to the pure holographic
dark energy model.

\section*{Acknowledgments}
XW would like to thank  Heng Yu and Xin Zhang for helpful
discussions. XW and ZHZ were supported by the National Natural
Science Foundation of China, under Grant No.10533010, 973 Program
No.2007CB815401 and Program for New Century Excellent Talents in
University (NCET) of China. RGC was supported in part by a grant
from Chinese Academy of Sciences (No. KJCX3-SYW-N2), and by NSFC
under grants No.~10325525, No.~10525060 and No.~90403029.


\begin{thebibliography}{99}

\bibitem{SN}
A.~G.~Riess {\it et al.} [Supernova Search Team Collaboration],
Astron.\ J.\  {\bf 116}, 1009 (1998) [astro-ph/9805201];\\
S.~Perlmutter {\it et al.} [Supernova Cosmology
ProjectCollaboration], Astrophys.\ J.\  {\bf 517}, 565 (1999)
[astro-ph/9812133].






\bibitem{WMAP}
D.~N.~Spergel {\it et al.} [WMAP Collaboration], [astro-ph/0603449];\\
L.~Page {\it et al.} [WMAP Collaboration], [astro-ph/0603450];\\
G.~Hinshaw {\it et al.} [WMAP Collaboration], [astro-ph/0603451];\\
N.~Jarosik {\it et al.} [WMAP Collaboration], [astro-ph/0603452].

\bibitem{cc}
P.~J.~E.~Peebles and B.~Ratra,
 Rev.\ Mod.\ Phys.\  {\bf 75}, 559 (2003) [astro-ph/0207347];\\
S.~M.~Carroll,
  Living Rev.\ Rel.\  {\bf 4}, 1  (2001)
  [astro-ph/0004075];\\
S.~Weinberg, Rev.\ Mod.\ Phys.\  {\bf 61}, 1 (1989).


\bibitem{0603057}
E.~J.~Copeland, M.~Sami and S.~Tsujikawa,
  Int.\ J.\ Mod.\ Phys.\  D {\bf 15}, 1753  (2006)
  [hep-th/0603057].

\bibitem{f(R)}
S.~Capozziello, S.~Carloni and A.~Troisi, [astro-ph/0303041];\\
S.~Capozziello, V.~F.~Cardone, S.~Carloni and A.~Troisi, Int.\ J.\ Mod.\ Phys.\ D {\bf 12}, 1969 (2003);\\
S.~M.~Carroll, V.~Duvvuri, M.~Trodden and M.~S.~Turner, Phys.\ Rev.\ D {\bf 70}, 043528 (2004);\\
R.~P.~ Woodard, astro-ph/0601672.


\bibitem{DGP}
G.~Dvali, G.~Gabadadze and M.~Porrati, Phys.\ Lett.\ B {\bf 485}, 208 (2000);\\
G.~R.~Dvali, G.~Gabadadze, M.~Kolanovic and F.~Nitti, Phys.\ Rev.\ D
{\bf 64}, 084004 (2001) ;\\
G.~R.~Dvali, G.~Gabadadze, M.~Kolanovic and F.~Nitti, Phys.\ Rev.\ D
{\bf 65}, 024031 (2002).

\bibitem{Li}
M.~Li, Phys.\ Lett.\  B {\bf 603}, 1 (2004) [hep-th/0403127].

\bibitem{cohen}
A.~Cohen, D.Kaplan and A.~Nelson, Phys.\ Rev.\ Lett. {\bf 82}, 4971
(1999); [hep-th/9803132]

\bibitem{Hsu}
S.~D.~H.~Hsu, hep-th/0403052

\bibitem{holo}
G.~'t Hooft, [gr-qc/9310026];\\
L.~ Susskind,  J. Math. Phys. 36, 6377  (1995) [hep-th/9409089]

\bibitem{RS}
L.~Randall and R.~Sundrum, Phys.\ Rev.\ Lett. {\bf 83}, 4690
(1999);\\
L.~Randall and R.~Sundrum, Phys.\ Rev.\ Lett. {\bf 83}, 3370 (1999).

\bibitem{LDGP}
V.~Sahni and Y.Shtanov, JCAP {\bf 0311} (2003) 014 [astro-ph/0202346];\\
A.~Lue and G.~D.~Starkman, Phys.\ Rev.\ D {\bf 70},
101501 (2004) [astro-ph/0408246];\\
R.~Lazkoz, R.~Maartens and E.~Majerotto,
  Phys.\ Rev.\ D {\bf 74}, 083510 (2006)
  [arXiv:astro-ph/0605701].

\bibitem{QDGP}
  L.~P.~Chimento, R.~Lazkoz, R.~Maartens and I.~Quiros,
  JCAP {\bf 0609}, 004 (2006)
  [arXiv:astro-ph/0605450].

\bibitem{SDGP}
Phys.\ Rev.\ D {\bf 75}, 023510 (2007)  [astro-ph/0611834].



\bibitem{CDGP}
M.~Bouhmadi-L\'{o}pez and R.~Lazkoz,  arXiv:0706.3896v1 [astro-ph]

\bibitem{holo fit}
Q.~G.~Huang and Y.~G.~Gong,
   JCAP {\bf 0408}, 006 (2004)
  [astro-ph/0403590];\\
X.~Zhang and F.~Q.~Wu,
   Phys.\ Rev.\  D {\bf 72}, 043524 (2005)
  [astro-ph/0506310];\\
Z.~Chang, F.~Q.~Wu and X.~Zhang,
  Phys.\ Lett.\  B {\bf 633}, 14 (2006)
  [astro-ph/0509531];\\
Z.~L.~Yi and T.~J.~Zhang,
  Mod.\ Phys.\ Lett.\  A {\bf 22}, 41 (2007)
  [astro-ph/0605596];\\
X.~Zhang and F-Q.~Wu, Phys.\ Rev.\ D {\bf 76}, 023502 (2007)
[astro-ph/0701405].

\bibitem{Davis}
Davis {\it et al.} 2007 [astro-ph/0701510].


\bibitem{nearby SNe}

M.~Hamuy, M.~M.~Phillips, N.~B.~Suntzeff, R.~A.~Schommer and
J.~Maza,
Astron.\ Jour.\ {\bf 112}, 2408  (1996)
  [astro-ph/9609064];\\
S.~Jha, A.~G.~Riess and R.~P.~Kirshner,
  Astrophys.\ J.\  {\bf 659}, 122 (2007)
  [astro-ph/0612666].

\bibitem{ESSENCE}
Wood-Vasey {\it et al.} 2007 [astro-ph/0701041].

\bibitem{SNLS}
  P.~Astier {\it et al.},
  Astron.\ Astrophys.\  {\bf 447}, 31 (2006)
  [arXiv:astro-ph/0510447].

\bibitem{HST}
A.~G.~Riess {\it et al.},
 Astrophys.\ J.\   {\bf 659}, 98  (2007)
[arXiv:astro-ph/0611572].

\bibitem{statistics}
S.~Nesseris and L.~Perivolaropoulos,
 Phys.\ Rev.\ D {\bf 72}, 123519  (2005) [astro-ph/0511040];\\
L.~Perivolaropoulos,
 Phys.\ Rev.\ D {\bf 71}, 063503 (2005) [astro-ph/0412308];\\
E.~Di Pietro and J.~F.~Claeskens,
 Mon.\ Not.\ Roy.\ Astron.\ Soc.\  {\bf 341}, 1299 (2003)
 [astro-ph/0207332].

\bibitem{BAO}
D.~J.~Eisenstein, {\it et al.} ApJ 633, 560 (2005)
[astro-ph/0501171].

\bibitem{SDSS}
M.~Tegmark {\it et al.} [SDSS Collaboration],  Phys.\ Rev.\ D {\bf 69}, 103501  (2004) [astro-ph/0310723];\\
M.~Tegmark {\it et al.} [SDSS Collaboration],  Astrophys.\ J.\  {\bf 606}, 702  (2004) [astro-ph/0310725];\\
U.~Seljak {\it et al.}, Phys.\ Rev.\ D {\bf 71}, 103515  (2005) [astro-ph/0407372];\\
J.~K.~Adelman-McCarthy {\it et al.}  [SDSS Collaboration],  Astrophys.\ J.\ Suppl.\  {\bf 162}, 38 (2006) [astro-ph/0507711];\\
K.~Abazajian {\it et al.} [SDSS Collaboration], [astro-ph/0410239]; [astro-ph/0403325]; [astro-ph/0305492];\\
M.~Tegmark {\it et al.} [SDSS Collaboration], Phys.\ Rev.\  D
{\bf74}, 123507 (2006) [astro-ph/0608632].



\bibitem{DGP test}
C. Deffayet, Lett. B \textbf{502}, 199 (2001);\\
D. Jain, A. Dev \& J. S. Alcaniz, Phys. Rev. D \textbf{66}, 083511
(2002);\\
J. S. Alcaniz, D. Jain \& A. Dev, Phys. Rev. D \textbf{66}, 067301
(2002);\\
J. S. Alcaniz \& Z. H. Zhu, Phys. Rev. D \textbf{71}, 083513
(2005);\\
Z. H. Zhu \& J. S. Alcaniz, ApJ. \textbf{620}, 7 (2005);\\
N. Pires, Z. H. Zhu \& J. S. Alcaniz, Phys. Rev. D \textbf{73},
123530 (2006);\\
Z. K. Guo, Z. H. Zhu, J. S. Alcaniz \& Y. Z. Zhang. ApJ.
\textbf{646}, 1 (2006);\\
R.~Maartens and E.Majerotto, Phys. Rev. D {\bf 74},  023004 (2006)
[astro-ph/0603353].

\end{thebibliography}
\end{document}